\documentclass[pra,twocolumn,showpacs,amsmath,amssymb]{revtex4-1}

\usepackage{graphicx}
\usepackage{amsfonts}
\usepackage{amsmath}
\usepackage{amssymb}
\usepackage{color}
\usepackage{bm}
\usepackage{bbm}
\usepackage{enumerate}
\usepackage{color}
\graphicspath{figures} 
\usepackage{amsfonts, amsmath, amsthm, amssymb} 
\usepackage{mathtools}
\usepackage{array}
\usepackage[colorlinks,bookmarks=false,citecolor=blue,linkcolor=red,urlcolor=blue]{hyperref}

\usepackage[colorlinks,bookmarks=false,citecolor=blue,linkcolor=red,urlcolor=blue]{hyperref}
\usepackage[usenames,dvipsnames,svgnames,table]{xcolor}
\graphicspath{figures} 

\newcommand{\be}{\begin{equation}}
\newcommand{\bee}{\begin{equation*}}
\newcommand{\ee}{\end{equation}}
\newcommand{\eee}{\end{equation*}}
\newcommand{\bearre}{\begin{eqnarray*}}
\newcommand{\eearre}{\end{eqnarray*}}
\newcommand{\bearr}{\begin{eqnarray}}
\newcommand{\eearr}{\end{eqnarray}}

\begin{document}

\title{
	Multi-speed prethermalization in spin models with power-law decaying interactions
}

\author{Ir\'en\'ee Fr\'erot$^1$
\footnote{Electronic address: \texttt{irenee.frerot@ens-lyon.fr}},
	Piero Naldesi $^{2,3}$
 and Tommaso Roscilde$^{1,4}$
 }

\affiliation{$^1$ Univ Lyon, Ens de Lyon, Univ Claude Bernard, CNRS, Laboratoire de Physique, F-69342 Lyon, France}
\affiliation{$^2$ Dipartimento di Fisica e Astronomia dell'Universit\`a di Bologna, Via Irnerio 46, 40127 Bologna, Italy}
\affiliation{$^3$ INFN, Sezione di Bologna, Via Irnerio 46, 40127 Bologna, Italy}
\affiliation{$^4$ Institut Universitaire de France, 103 boulevard Saint-Michel, 75005 Paris, France}
\date{\today}


\begin{abstract}
 The relaxation of uniform quantum systems with finite-range interactions after a quench is generically driven by the ballistic propagation of long-lived quasi-particle excitations triggered by a sufficiently small quench. Here we investigate the case of long-range ($1/r^{\alpha}$) interactions for $d$-dimensional lattice spin models with uniaxial symmetry, and show that, in the regime $d < \alpha < d+2$, the entanglement and correlation buildup is radically altered by the existence of a \emph{non-linear} dispersion relation of quasi-particles, $\omega\sim k^{z<1}$, at small wave vectors, leading to a divergence of the quasiparticle group velocity and \emph{super}-ballistic propagation. This translates in a super-linear growth of correlation fronts with time, and sub-linear growth of relaxation times of subsystem observables with size, when focusing on $k=0$ fluctuations. Yet the large dispersion in group velocities leads to an extreme wavelength dependence of relaxation times of finite-$k$ fluctuations, with entanglement being susceptible to the longest of them. Our predictions are directly relevant to current experiments probing the nonequilibrium dynamics of trapped ions, or ultracold magnetic and Rydberg atoms in optical lattices.   
\end{abstract}

\maketitle

\begin{figure}
 \includegraphics[width=0.9\linewidth]{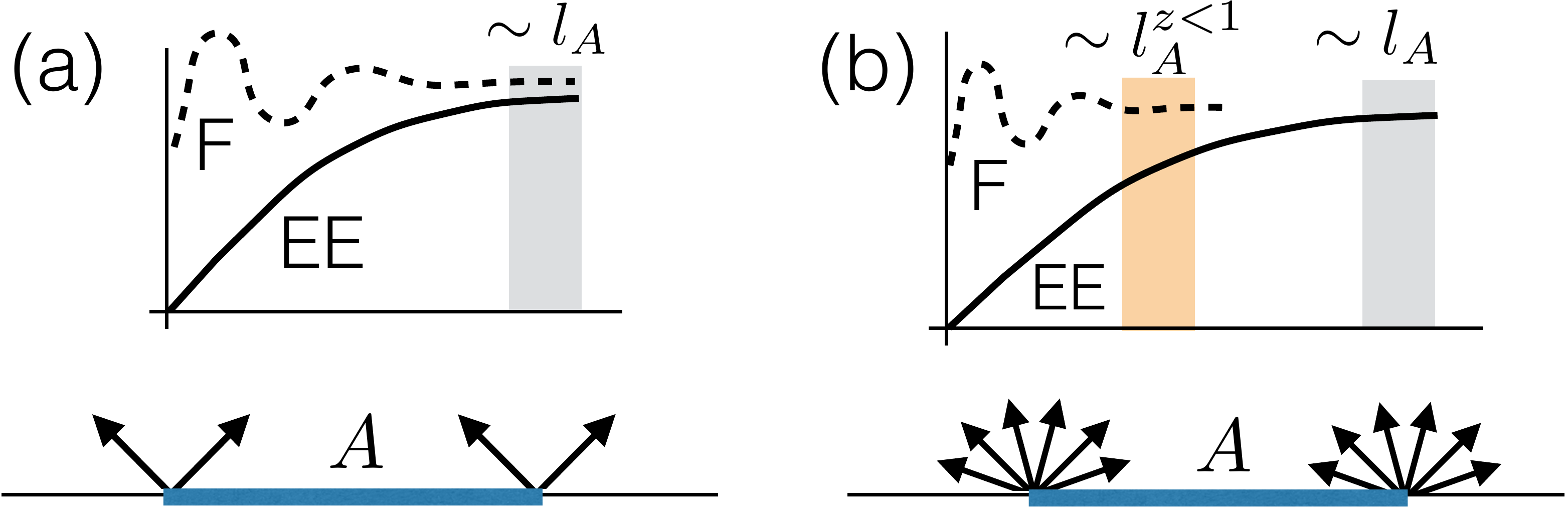}
 \caption{Pre-thermalization scenario for short- vs. long-range interactions. (a) For SR interactions, the prethermalization process in a subsystem ($A$) exhibits a unique characteristic time scale for both entanglement entropy (EE) and fluctuations (F), given by the ratio of the linear size of the subsystem ($l_A$) to twice the maximum group velocity of the quasiparticle excitations \cite{CalabreseC2016}; (b) For LR interactions, the QP group velocity may range from infinity to zero, introducing a broad range of time scales: the dynamics of long-wavelength fluctuations is sensitive to the fastest modes, while entanglement dynamics is sensitive to all modes including the slowest ones.}
 \label{fig_schema}
\end{figure}

\textit{Introduction.}  The relaxation of the pure state of a generic quantum many-body system following a quantum quench is a complex phenomenon: there each subsystem sees its reduced state reach thermal equilibrium (or a non-equilibrium steady-state) via the drastic rearrangement of correlations with its exterior, acting as a bath \cite{lindenetal2009, dalessionetal2016}. A particularly dramatic case -- yet highly relevant experimentally -- concerns the choice of a factorized initial state, in which each subsystem admits a pure-state (zero-entropy) description. The buildup of entropy and fluctuations towards equilibration requires therefore the growth of entanglement and correlations between subsystems, namely the flow of information from each subsystem to its exterior. The post-quench dynamics of entanglement and correlations strongly depends \emph{a priori} on the peculiar system under investigation. Yet, in generic systems with short-range (SR) interactions and translational invariance, a seemingly universal picture emerges for sufficiently small quenches, as shown both theoretically (see Ref.~\cite{CalabreseC2016} and refs therein) as well as experimentally \cite{cheneauetal2012, Langen2013}. In such a picture (see Fig.~\ref{fig_schema}(a) for a sketch) correlations and entanglement establish because of the ballistic propagation of quasiparticle excitations (QP) triggered by the quench. As a consequence 1) correlations reorganize over distances which grow linearly in time (within a so-called ``causal cone" or ``light cone") consistent with Lieb-Robinson bounds \cite{liebR1972} on the propagation of signals in quantum systems -- and, as a consequence, subsystem fluctuations (which are integrals of correlation functions) grow linearly in time as well. The velocity of correlation fronts is twice the maximum group velocity of the QP -- this is consistent with correlations among two points being established by counter-propagating QP originating from the midpoint; 2) entanglement entropy (EE) grows linearly in time as a consequence of ballistic QP traversing a subsystem \cite{calabreseC2005}.        

 Very interestingly, the above universal picture of post-quench dynamics breaks down in the presence of long-range (LR) interactions which decay as a power-law $1/r^{\alpha}$ of the distance $r$ -- stirring a considerable recent activity both in theory \cite{haukeT2013, schachenmayeretal2013, eisertetal2013, hazzardetal2014, foss-feigetal2015, schachenmayeretal2015, cevolanietal2015, buyskikhetal2016, maghrebietal2016} and experiments \cite{richermeetal2014, Jurcevic2014}. In particular, in contrast with SR interactions, one generally observes that 1) correlation fronts can move in a super-ballistic fashion (namely faster than linearly in time) for sufficiently small values of $\alpha$; 2) the growth of EE can be sub-linear in time (and as slow as logarithmic for small $\alpha$). This very rich phenomenology lacks universal traits, and it escapes so far a unifying picture; as an example, rigorous generalizations of the Lieb-Robinson bounds to LR interactions \cite{Hastings2006,Gongetal2014,foss-feigetal2015} reveal to be loose compared to the actual observations on specific systems. At the same time the importance of closed, nonequilibrium LR-interacting systems is augmented by their experimental realization in atomic physics via trapped ions \cite{richermeetal2014, Jurcevic2014}, magnetic atoms \cite{dePazetal2013}, Rydberg atoms \cite{Browaeysetal2016} and dipolar molecules \cite{Moses2017}.

 Inspired by the potential of the above cited experimental setups to realize quantum spin models with LR interactions, we provide in this work a complete picture of the quench dynamics in paradigmatic XX spin models with $1/r^{\alpha}$ interactions in arbitrary ($d$) dimensions. Indeed we show that a QP description -- via linear spin-wave (LSW) theory, corroborated in $d=1$ by density-matrix renormalization group (DMRG) calculations -- allows one to achieve a unifying picture for the dynamics of correlations, fluctuations and entanglement. The contradiction between the fast dynamics of correlations and fluctuations and the slow growth of entanglement is resolved by a scenario of \emph{multi-speed prethermalization} \footnote{We use the term \emph{prethermalization} here as our QP treatment of the system can only describe dephasing between independent modes, while true thermalization is achieved over longer time scales due to the non-linear coupling terms among QP modes.} (see sketch in Fig.~\ref{fig_schema}(b)), in which LR interactions lead to the divergence of the maximum group velocity, so that the velocities of QPs moving with different wavevectors span an infinite range. Different Fourier components of local fluctuations possess relaxation times which scale as the inverse of the corresponding group velocities, allowing one to reconstruct (both theoretically as well as experimentally) the multi-speed nature of the prethermalization process. The growth of the EE, on the other hand, is sensitive to the dynamics of fluctuations at all wavelengths, and as a consequence it is much slower than the evolution of the fastest fluctuations.      
 
\textit{Model and methods.---}
We center our analysis on the quantum XX model with $1/r^{\alpha}$ interactions (hereafter denoted as $\alpha$XX model), describing quantum $s$-spins localized at the nodes of a $d-$dimensional lattice, interacting with a power-law decaying exchange interaction : 
\be 
	{\cal H}_{\alpha \rm XX} = -J \sum_{i\neq j} \vert {\bm r_i} - {\bm r_j} \vert^{-\alpha} \left ( S_i^x S_j^x + S_i^y S_j^y \right ) ~.
\label{H_XX}
\ee
Here $S_i^{\beta}$ ($\beta = x,y,z$) are quantum spin operators defined on the sites of a $d$-dimensional hypercubic lattice. The results presented here are restricted to the case $s=1/2$, although they are immediately generalizable to arbitrary $s$ values. 
Special emphasis is put on the case $\alpha=3$, relevant to dipolar systems \cite{dePazetal2013, hazzardetal2014, Browaeysetal2016}. We focus our study on the unitary quench dynamics $|\Psi(t)\rangle = \exp(-i{\cal H}_{\alpha\rm XX}t) |\Psi(0)\rangle$ generated by the $\alpha$XX Hamiltonian starting from the factorized state $|\Psi(0)\rangle = |\uparrow_x\rangle^{\otimes N}$ of spins polarized along the $x$ axis, which is both of fundamental interest, as well as of practical relevance for experiments \footnote{In the DMRG calculations we consider the above state projected onto the $S_{\rm tot}^z=0$ subspace, in order to decrease the size of the Hilbert space visited by the evolved state.}. When considering the case of ferromagnetic interactions $J>0$, one immediately recognizes that $|\Psi(0)\rangle$ is the mean-field ground state of the Hamiltonian -- which is expected to be an increasingly good approximation to the exact ground state of the $\alpha$XX model for large $d$ and small $\alpha$ \cite{paper1-XX_LR}. As a consequence, the energy injected by the quench leads the system to relax towards a low-temperature regime which exhibits long-range ferromagnetic order in the $xy$ plane -- this is true in $d=3$ for any $\alpha$, while it requires $\alpha < 2d$ in $d=1,2$ to stabilize long-range order at low but finite temperatures \cite{SuppMat} \footnote{In fact in $d=2$ even for $4 < \alpha < \infty$ the system exhibits quasi-long-range order at low temperature, so that our treatment remains justified.}. Under these circumstances, a description of the Hamiltonian in terms of harmonic fluctuations around the mean-field ground state -- namely LSW theory -- is very well justified, and its quantitative validity can be monitored systematically during the time evolution. Moreover, as discussed in the Supplementary Material (SM - \cite{SuppMat}) the time-reversal symmetry of the $\alpha$XX Hamiltonian and of the initial state leads to the remarkable property that the forward- and backward-evolved states $|\Psi(t)\rangle$ and $|\Psi(-t)\rangle$ give the same expectation values for time-reversal symmetric observables as well as for the EE: as a consequence all of our results apply as well to the evolution driven by the \emph{antiferromagnetic} Hamiltonian ($J<0$), for which the initial state we choose is a superposition of highly excited eigenstates. 

\begin{figure}[t]
 \includegraphics*[width=0.9\linewidth]{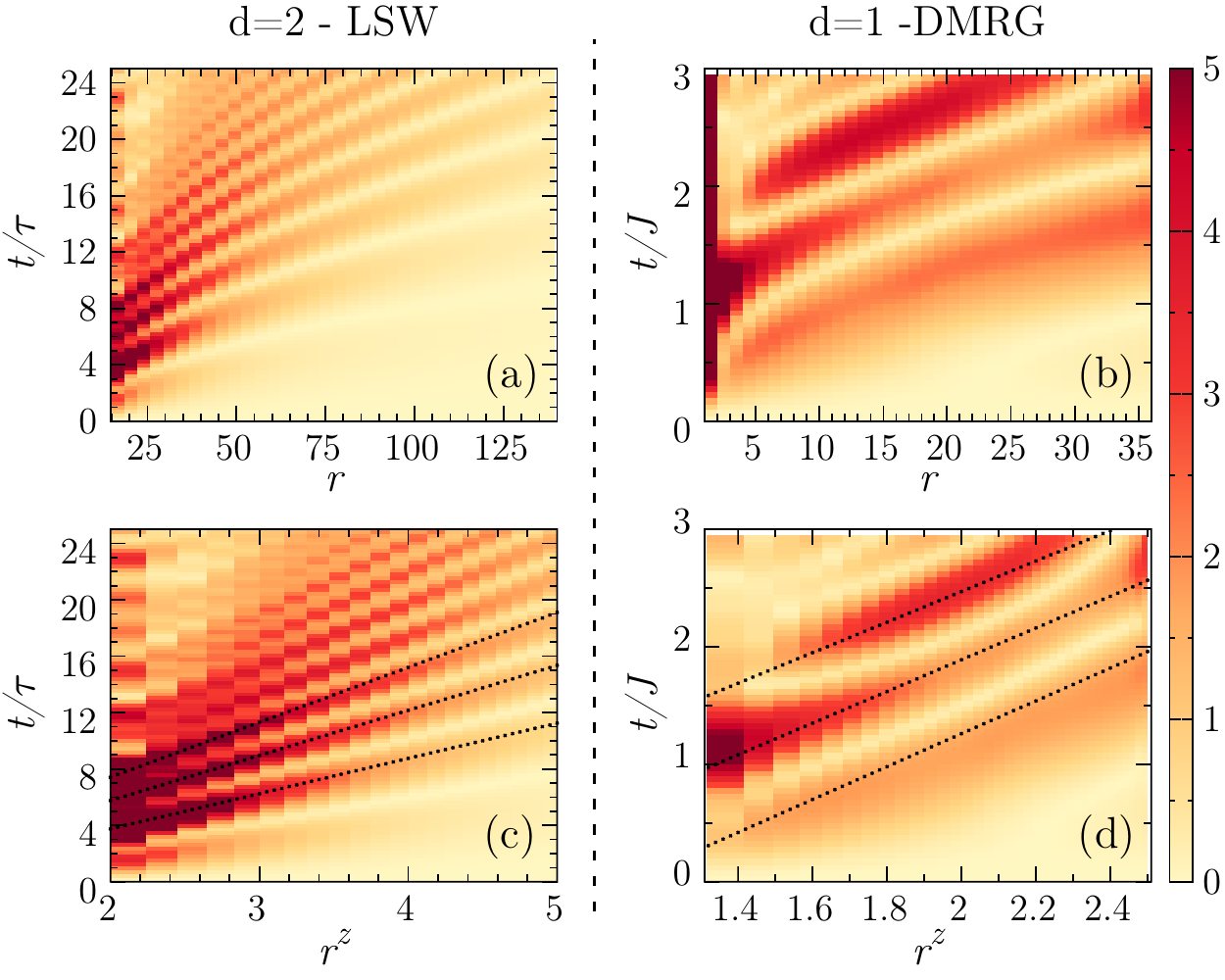}
  \caption{Curved light-cone dynamics. (a-b) Evolution of the correlation function $\vert C^{zz}({\bm r},t) \vert$ (a) $\alpha=3$ in $d=2$ from LSW theory ($\bm r = (r,0)$ and $L_x=L_y=300$, the signal has been multiplied by $1000$; time is in units $\tau = (2sJ\gamma_0)^{-1}$); and (b) $\alpha=1.5$ in $d=1$ from DMRG ($L=40$, signal multiplied by $50$). (c-d) Same as (a-b) but plotted in the $(r^{z},t)$ plane, where $z=1/2$ in (c) and $z=1/4$ in (d). Dotted lines are guides to the eye.} 
 \label{f.lightcone}
\end{figure}

 The central output of LSW theory \cite{paper1-XX_LR, SuppMat} is the reduction of the Hamiltonian Eq.~\eqref{H_XX} to that of free bosonic QP, ${\cal H} \approx \sum_{\bm k} \omega_{\bm k} \beta_{\bm k}^{\dagger} \beta_{\bm k}$, with the dispersion relation 
\be 
		\omega_{\bm k} = 2sJ \gamma_0 \sqrt{1 - \gamma_k/\gamma_0} ~
	\label{eq_omega_k}
	\ee
	where  $\gamma_k = \sum_{\bm r \neq 0} e^{i{\bm k\cdot \bm r}} |{\bm r}|^{-\alpha}$ is the Fourier transform of the interaction. The long-wavelength behavior of this dispersion relation is strongly affected by the value of the $\alpha$ exponent: indeed for small $k$ one finds \cite{paper1-XX_LR} that $\omega_{\bm k} \approx c(\alpha) k^z$, where the exponent $z$ has value $z=1$ for $\alpha \geq d+2$ -- continuously connecting with the well-known case of nearest-neighbor interactions \cite{Mattis-book} -- while $z = (\alpha-d)/2 < 1$ for $d \leq \alpha \leq d+2$, and $z=0$ for $\alpha \leq d$. When $z=1$ the QP excitations have a well-defined, $\alpha$-dependent maximum group velocity $v_{\rm max} = c$, which also sets a characteristic time scale $t^*(l) = l/(2c)$ for relaxation of observables in a subsystem of linear size $l$, in accordance with the universal SR picture of post-quench dynamics. On the other hand, the regime with $z<1$ is characterized by a \emph{divergent} group velocity as $k\to 0$, $v_{\rm max} \sim k^{z-1}$, suggesting the breakdown of the SR picture. This result is on par with the field-theoretical (\emph{i.e.} long-wavelength) description of the Hamiltonian Eq.~\ref{H_XX}, corresponding to a 2-component $(d+1)$-dimensional critical field theory with dynamical critical exponent $z$ \cite{SuppMat}. In the following we shall particularly focus on the intermediate regime $d \leq \alpha \leq d+2$, exemplified by the experimentally relevant case of dipolar interactions ($\alpha=3$) in $d=2$. \footnote{The regime $\alpha\leq d$ is continuously connected to the case of infinite-range interactions, describing the dynamics of the collective spin -- we shall postpone the study of this regime to future work.}
	
 \emph{From straight to curved light cones.} A fundamental trait of the post-quench dynamics is represented by the characteristic buildup of spin-spin correlations starting from the uncorrelated, factorized state  -- in the following we shall concentrate on the correlation function for the $z$ spin component $C^{zz}(\bm r,t) = \langle \delta S_i^z \delta S_{i+\bm r}^z \rangle (t)$ where $\delta O = O - \langle O \rangle$; (see \cite{SuppMat} for the details of its LSW calculation). In the case of $\alpha > d+2$ (so that $z=1$) our LSW results in $d=2$, as well as the DMRG ones in $d=1$, show a ballistic propagation of the correlation front -- in the LSW calculation the speed of propagation is consistent with 2$c(\alpha)$. When instead the dynamical exponent $z$ becomes different from 1, the long-wavelength dynamics of the system loses the Lorentz invariance characteristic of the $z=1$ case, and time and space enter very differently in the low-energy description. This insight suggests that the distance over which correlations establish at time $t$ advances as $r \sim t^{1/z}$, namely \emph{super}-ballistically, as a result of the divergence of the group velocity of the fastest QP. This is precisely what is observed in both the $d=2$ LSW data [see Fig. \ref{f.lightcone}(a,c)] and the $d=1$ DMRG data [Fig.~\ref{f.lightcone}(b,d)]: the space-time correlation fronts acquire a finite curvature which is quantitatively explained by the non-linear relationship between space and time established by the LR interactions. The centrality of the $z$ exponent in determining the shape of correlation fronts has been very recently established for the linear response to local quenches in the field theoretical description of the critical $1d$ Ising model with LR interactions, and of $1d$ free fermions with LR hopping and pairing \cite{maghrebietal2016}. Our results extend the role of the $z$ exponent to the case of global quenches beyond linear response, and within the microscopic description of bosonic critical and LR-interacting models.
 
 \begin{figure*}[t]
 \includegraphics*[width=\linewidth]{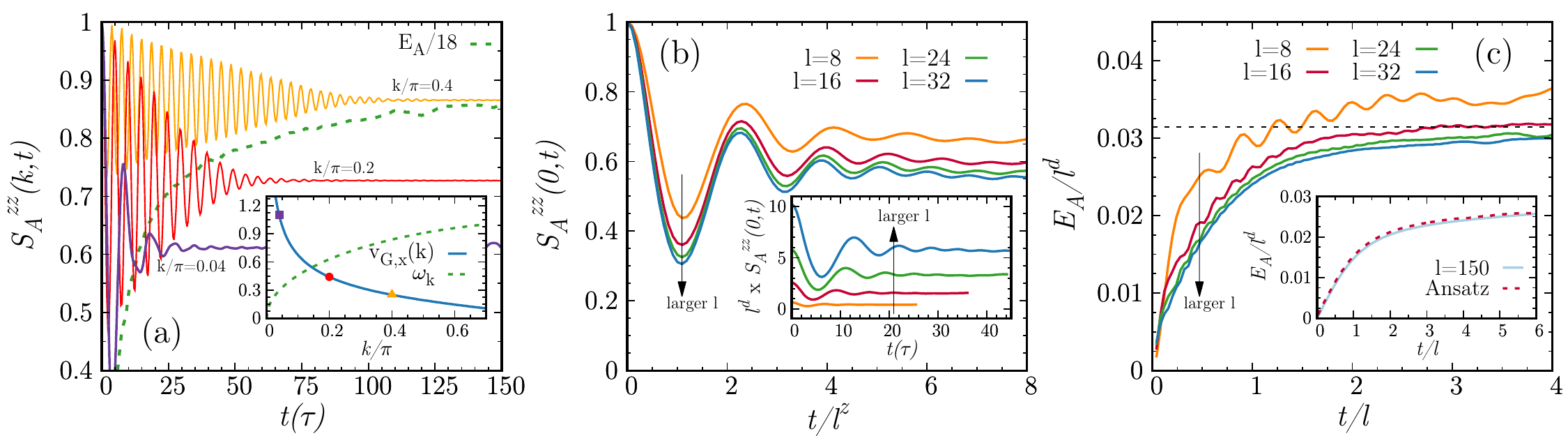}
  \caption{\emph{Multispeed prethermalization of a subsystem}. (a) Evolution of the subsystem structure factor $S_A^{zz}(\bm k,t)$ for different wavevectors $\bm k = (k,0)$ on a $50 \times 10$ rectangle immersed in a 2$d$ $400 \times 400$ lattice with $\alpha =3$. Dashed line: evolution of EE of the same subsystem. Inset: corresponding frequency and group velocity along the $(k,0)$ axis. Symbols mark the wavevectors at which the structure factor is evaluated on the main panel. Frequency in units of $2sJ\gamma_0$; (b) Evolution of the magnetization fluctuations $\langle \delta^2 S_A^z \rangle(t)/l^d = S_A^{zz}(0,t)$ for different subsystem sizes. Region $A$ is a $l \times l$ square in a $10l \times 10l$ torus for $l=8,16,24,32$. Time is rescaled to $l^z$. Inset: same as main panel without rescalings; (c) Evolution of the entanglement entropy $E_A(t)/l^d$ as a function of the rescaled time $t/l$-- same geometry and sizes as in (b). Horizontal dashed line is the stationary GGE prediction (see text). Inset: evolution of EE for a $150 \times 150$ cylinder in a $150 \times 1500$ torus. Dashed line is the asymptotic prediction of Eq. \eqref{e.Sconjecture}. In all panels, time is in units of $\tau=(2sJ\gamma_0)^{-1}$.}
 \label{f.multispeed}
\end{figure*}	
 
  \emph{Multi-speed evolution of finite-$k$ fluctuations.} The appearance of a curved space-time correlation front, while revealing the existence of a divergent group velocity for QP, remains silent about the fact that this group velocity spans the whole interval from $\infty$ to 0 when moving from the center to the edge of the Brillouin zone. But a deeper analysis of the spatial structure of correlations \emph{inside} the curved light cone can capture the extreme dispersion of group velocities of QPs. The guiding idea is that the group velocity governs the evolution of \emph{wave packets} -- namely of superpositions of Fourier modes with frequencies tightly distributed around a reference value. For the quench under examination, LSW predicts that the evolution of the spin structure factor $S^{zz}(\bm k) = N^{-1} \sum_{ij\in N} e^{i{\bm k}\cdot(\bm r_i - \bm r_j)} \langle \delta S_i^z \delta S_j^z \rangle_t$ contains a constant part plus a term oscillating at frequency $2\omega_k$ 
  \begin{equation}
  S^{zz}(\bm k,t) =  g^{(0)}_{\bm k} +  g^{(1)}_{\bm k}\cos (2\omega_{\bm k} t)
  \end{equation}
  where $g^{(1)}_{\bm k} = s \gamma_{\bm k}/4\gamma_0$ and $g^{(0)}_{\bm k} = s/2-g^{(1)}_{\bm k}$. In order to superpose these Fourier modes into a ``wave packet" one can consider the evolution of the structure factor on a subsystem $A$ of linear size $l_A$ (obtained by replacing $N^{-1}\sum_{ij\in N} \to l_A^{-d} \sum_{ij\in A}$ in the above expression), which takes the form of a convolution sum in Fourier space
 \begin{equation}
 S_A^{zz}(\bm k,t) = l_A^{-d}\sum_{\bm q} f_A(\bm q) S^{zz}(\bm k+\bm q,t)~. 
 \end{equation} 
 Here $f_A$ is the Fourier transform of the characteristic function of subsystem $A$ \cite{SuppMat}, which, for $l_A \gg 1$, has support roughly corresponding to the tight interval $[-\pi/l_A,\pi/l_A]^{\otimes d}$. As a result the subsystem structure factor takes the approximate form 
 \begin{equation}
 S_A^{zz}(\bm k,t) \approx g^{(0)}_{\bm k} + g^{(1)}_{\bm k} \cos(2\omega_{\bm k} t) \prod_{a=1}^d {\rm sinc} \left(\pi t/ t^*_{\bm k,a} \right)
 \label{e.SzzA}
 \end{equation}
 where $t_{\bm k,a}^* = l_A/[2 v_{G,a}(\bm k)]$ and $v_{G,a}(\bm k) = \partial_{k_a} \omega_{\bm k}$ is the $a$-th component of the group velocity, and ${\rm sinc}(x) = \sin (x)/x$.  Hence the subsystem structure factor exhibits the characteristic time dependence of a wave packet, with oscillations at frequency $2\omega_{\bm k}$ contained within an envelope that decays over a characteristic time scale $t_{\bm k}^* = \min_{a=1\dots d} t_{\bm k,a}^*$. Otherwise stated, $S_A^{zz}(\bm k,t)$ is a sum of close-by frequencies which dephase over a characteristic time scale $t_{\bm k}^*$, directly related to the group velocity. As shown in the SM \cite{SuppMat}, Eq.~\eqref{e.SzzA} provides an excellent approximation to the LSW data for $l_A\sim {\cal O}(10)$. This implies that the subsystem structure factor exhibits a peculiar form of multispeed pre-thermalization, in which each Fourier component relaxes over a strongly ${\bm k}$-dependent time scale, which in turn is linear in the subsystem size for finite $\bm k$ -- see Fig.~\ref{f.multispeed}(a). On the other hand for $k = 0$ the group velocity diverges, which for a subsystem of size $l_A$ implies that $v_{G}(0) \sim c ~l_A^{1-z}$. As a consequence the $k=0$ component of the structure factor, related to the fluctuations of the uniform subsystem magnetization as $\langle \delta^2 S_A^z\rangle/l_A^d$, relaxes over a characteristic time scale  $t_{0}^* \sim l_A^z/(2c)$, scaling \emph{sub-linearly} with subsystem size. This is consistent with the super-ballistic propagation of correlations -- of which $\langle \delta^2 S_A^z\rangle$ simply represents the integral over a subsystem --  and it is clearly exemplified by the data of Fig.~\ref{f.multispeed}(b) for a 2$d$ system with dipolar interactions. Furthermore, following this line of thinking one can introduce the Fourier transform of the spin on a subsystem $A$, $S^z_{\bm k,A} = l_{A}^{-d/2} \sum_{i \in A} e^{i{\bm k}\cdot {\bm r}_i} S_i^z$, and the ${\bm k}$-dependent correlation function $C^{zz}_{\bm k}(A_i, A_j) = \langle \delta S^z_{\bm k,A_i} \delta S^z_{-\bm k,A_j} \rangle $, where $A_i$ and $A_j$ are two subsystems at distance ${\bm r}_{ij}$ -- so that $S_A^{zz}(\bm k) = \sum_{A_i,A_j \in A} C^{zz}_{\bm k}(A_i, A_j)$, where subsystems $A_i$'s tile regularly (namely without overlaps) the subsystem $A$.  As shown in the SM \cite{SuppMat}, the $\bm k$-dependent correlation function $C^{zz}_{\bm k}$ after the quench exhibits a correlation front which advances at a speed $2v_{G,a}(\bm k)$ in the $a$-th direction (or a \emph{${\bm k}$-dependent light cone}, by virtue of the strong group velocity dispersion), and consistently its integral on a region of linear size $l_A$ relaxes on the above-cited characteristic time scales $t_{\bm k}^*$.   
 
 \emph{Entanglement dynamics.} Finally we turn to the dynamics of the entanglement entropy (EE), defined as $E_A = - {\rm Tr} \rho_A \log \rho_A$, where $\rho_A$ is the reduced density matrix of subsystem $A$. The EE is sensitive to the relaxation of all forms of fluctuations, and therefore, due to the extreme dispersion in relaxation timescales for the Fourier components of the fluctuations, it can be expected to exhibit a much slower relaxation than long-wavelength fluctuations. Fig.~\ref{f.multispeed}(c) shows indeed that the EE obeys a scaling form $E_A(t) = l_A^d f_E(t/l_A)$, exhibiting a characteristic time scale which grows linearly with $l_A$; its detailed time dependence shows a fast, sublinear rise at short times, followed by significant slowdown resulting in a very broad shoulder (the broader the smaller $\alpha$). The EE dynamics can be quantitatively understood by conjecturing that it stems from the relaxation of independent QP modes, each contributing an entropy $s_{\rm eq}(\bm k) = (1+n_{\bm k}) \log (1+n_{\bm k}) - n_{\bm k} \log n_{\bm k}$ in the stationary state (generalized Gibbs ensemble, GGE \cite{VidmarR2016}), where $n_{\bm k} = \langle \beta^\dagger_{\bm k} \beta_{\bm k}\rangle$ is the conserved number of QP in each mode fixed by the initial state. One can then postulate the following form   
 \begin{equation}
 E_A(t) = \sum_{\bm k} f(t/t^*_{\bm k}) s_{\rm eq}(\bm k)
 \label{e.Sconjecture}
 \end{equation} 
where the sum is restricted to modes ${\bm k}$ whose wavelength is commensurate with subsystem $A$ (a similar form can be proved to be exact for quenches in one-dimensional free fermions \cite{fagottiC2008}). A simple choice of the $f$ function, $f(x) = \min(x,1)$, produces a very convincing agreement with the LSW data, as shown in Fig.~\ref{f.multispeed}(c). This result finally reconciles the slow EE dynamics with the ultrafast dynamics of correlations and long-wavelength fluctuations in LR interacting systems, by exposing the dependence of the EE on the slow evolution of short-wavelength fluctuations. 
 
 \emph{Conclusions and outlook.} Based on linear spin-wave and density-matrix renormalization group calculations, we have unveiled a scenario of multi-speed prethermalization for lattice spin models with long-range (LR) interactions. Our results reconcile the seemingly contradictory aspects of very fast evolution of correlations and long-wavelength fluctuations on a subsystem on the one hand, and of the very slow buildup of entanglement entropies on the other hand. This multi-speed scenario, obtained explicitly for the LR interacting XX model of magnetism, can be easily translated to other models of LR interactions studied in the recent past (especially the transverse field Ising model \cite{haukeT2013, schachenmayeretal2013, eisertetal2013, hazzardetal2014, foss-feigetal2015, schachenmayeretal2015, cevolanietal2015, maghrebietal2016}), allowing one to reconcile the discrepancy between the correlation dynamics and entanglement dynamics already observed in those systems. In order to reconstruct the variety of characteristic time scales governing the dynamics we offer a detailed diagnostic tool (namely the analysis of the subsystem structure factor) -- this is of immediate interest to current experiments on quantum gas microscopes, based on trapped ions \cite{richermeetal2014, Jurcevic2014} or neutral atoms \cite{cheneauetal2012,Labuhnetal2016}, which have direct access to the diagnostics in question. As an example, the multi-speed dynamics is realized by dipolar interactions in $d=2$, which are implemented by magnetic atoms \cite{dePazetal2013} and resonantly interacting Rydberg atoms \cite{Browaeysetal2016}. Hence our results pave the way for the reconstruction of most salient features of LR interaction dynamics to be searched in new-generation microscope experiments, as well as in systems beyond atomic physics (such as nano-engineered magnets \cite{Khajetooriansetal2015}, ensembles of nitrogen-vacancy centers \cite{Kucskoetal2016} and nuclear spins \cite{Alvarezetal2015}).
 
\textit{Acknowledgments} We thank A. Browaeys, G. Carleo, and M. Kastner for fruitful discussions and  M. Fagotti for correspondence. This work is supported by ANR (``ArtiQ" project).

\bibliographystyle{unsrt}
\bibliography{biblio_LR_dyn.bib}

\newpage
\appendix

\begin{center}
{\bf SUPPLEMENTARY MATERIAL for} \\
{\bf \emph{Multi-speed prethermalization in spin models with power-law decaying interactions}}
\end{center}

\section{Theorem on time-reversal symmetry of unitary evolution}

\subsection{General theorem}
In this section we prove the following general 
 
\underline{\emph{Theorem}.} Be ${\cal H}$ an Hamiltonian admitting a representation as a real-valued matrix on a given basis $|\phi\rangle$ of Hilbert space, and $|\Psi(0)\rangle$ an initial state which has all real coefficients when decomposed on the same basis. Then, for all observables $A$ which are represented as real-valued matrices on the basis, the evolution of the expectation value $\langle A \rangle(t) = \langle \Psi(t) | A | \Psi(t) \rangle$ is invariant under time reversal, namely
\begin{equation}
\langle A \rangle(t) =  \langle A \rangle(-t)~.
\end{equation}
Moreover, if the basis $|\phi\rangle$ is factorized among subsystems $A$ and $B$, $|\phi\rangle = |\phi^{(A)}\rangle \otimes |\phi^{(B)}\rangle$, then
\begin{equation}
S_A(t) = S_A(-t)
\end{equation}
where $S_A$ is the entanglement entropy of subsystem $A$. 

\medskip

\underline{\emph{Proof.}} The proof of the above result is readily achieved by inserting resolutions of the identities on the basis $|\psi_a\rangle$ of eigenstates of ${\cal H}$ with eigenenergies $E_a$:
\begin{align}
\langle A \rangle(t) =
\sum_{ab} \langle \psi_a | A |\psi_b\rangle \Psi_a \Psi_b  \left [\cos(\omega_{ab}t) + i \sin(\omega_{ab}t) \right ]
\label{e.At}
\end{align}
where we have introduced the frequencies $\omega_{ab} = (E_a - E_b)/\hbar$, and the decomposition of the initial state on the Hamiltonian eigenbasis:
\begin{equation}
|\Psi(0)\rangle = \sum_a \Psi_a |\psi_a\rangle 
\end{equation}
with $\Psi_a = \langle \psi_a | \Psi(0)\rangle$. 
The matrix elements and scalar products entering in Eq.~\eqref{e.At} are real by hypothesis (because they can always be reduced to products of real coefficients by inserting resolutions of the identities on the $|\phi\rangle$ basis). Hence the product $\langle \psi_a | A |\psi_b\rangle \langle \Psi_a \Psi_b \rangle$ is symmetric under the exchange of the $a$ and $b$ indices, and therefore the $\sin(\omega_{ab}t)$ term vanishes, leaving out only the term $\cos(\omega_{ab}t)$ term, which is even in $t$.

Given that $|\Psi(t)\rangle = \sum_a \Psi_a(t) |\psi_a\rangle$ with 
$\Psi_a(t) = \exp(-iE_at/\hbar) \Psi_a$ and $\Psi_a \in \mathbb{R}$, one has $\Psi_a(-t) = \Psi^*_a(t)$ -- namely time reversal amounts to simple conjugation of the coefficients of the decomposition on the Hamiltonian eigenbasis, or, equivalently, to conjugation of the coefficients of the decomposition on the basis $|\phi\rangle$. 
Consider then the Schmidt decomposition of the evolved state
 \begin{equation}
 |\Psi(t)\rangle = \sum_{p} \lambda_p(t) |a_p(t)\rangle \otimes |b_p(t)\rangle 
 \end{equation}
where $\lambda_p \in \mathbb{R}$. If the basis $|\phi\rangle$ admits the factorized form $|\phi_A\rangle \otimes |\phi_B\rangle$, then the basis of the Schmidt decomposition $|a_p(t)\rangle$ (resp. $|b_p(t)\rangle$) can be decomposed on the basis $|\phi_A\rangle$ (resp. $|\phi_B\rangle$) as  $|a_p(t)\rangle = \sum_l \varphi^{(A)}_{p,l}(t) |\phi_l^{(A)}\rangle$ (resp. $|b_p(t)\rangle = \sum_l \varphi^{(B)}_{p,l}(t) |\phi_l^{(B)}\rangle$). 
The fact that $|\Psi(-t)\rangle$ is related to $|\Psi(t)\rangle$ by conjugation of the coefficients of the decomposition on the  $|\phi\rangle$ basis implies that $\varphi^{(A)}_{p,l}(-t) = (\varphi^{(A)}_{p,l})^*(t)$ and $\varphi^{(B)}_{p,l}(-t) = (\varphi^{(B)}_{p,l})^*(t)$, but that $\lambda_p(-t) = \lambda_p(t)$. Given that the entanglement entropy only depends on the Schmidt coefficients $\lambda_p$ as $S_A = - \sum_p \lambda_p \log \lambda_p$, we conclude that it is invariant under time reversal. 
 
\subsection{Discussion}

The hypotheses of the above theorem are satisfied by the initial state $|\Psi(0)\rangle$ we choose in this work, as well as by the  $\alpha$XX Hamiltonian ${\cal H}_{\alpha{\rm XX}}$, due to its time-reversal symmetry - the basis $|\phi\rangle$ is just provided by the computational basis of eigenstates of the $S_i^z$ operators. All the observables of interest to this work - namely correlation functions - also satisfy the hypothesis of real valuedness on the same basis. Finally the computational basis is fully factorized, namely it is factorized for any spatial $A$-$B$ bipartition. 

The above result amounts to say that the Hamiltonians ${\cal H}_{\alpha{\rm XX}}$ and $-{\cal H}_{\alpha{\rm XX}}$ generate the same dynamics for the observables of interest to this work starting from the chosen state $|\Psi(0)\rangle$. This is a remarkable fact, given that the two Hamiltonians have quite different low-energy properties: ${\cal H}_{\alpha{\rm XX}}$ is ferromagnetic and not frustrated, while $-{\cal H}_{\alpha{\rm XX}}$ is antiferromagnetic and strongly frustrated due to the long-range nature of the couplings. 

\section{Details on the spin-wave approximation}
In this section we provide additional details on the semi-classical approximation used to study the dynamics of the $\alpha$-XX model of Eq. \eqref{H_XX} of the main text. 

\subsection{Assumptions behind spin-wave theory}

The quench dynamics we studied starts from the mean-field ground state of the $\alpha$-XX model, $|\Psi(0)\rangle = |\uparrow_x\rangle^{\otimes N}$. Under the mean-field approximation $S_i^{\beta}S_j^{\beta} \approx S_i^{\beta}\langle S_j^{\beta} \rangle + \langle S_i^{\beta}\rangle S_j^{\beta} - \langle S_i^{\beta}\rangle \langle S_j^{\beta} \rangle$, the chosen initial state does not evolve, because $\langle S_j^{y} \rangle = 0$, so that the mean-field Hamiltonian is proportional to the $S_i^x$ operator.  
 Our semi-classical treatment of the ensuing dynamics (based on linear spin-wave - LSW - theory) goes beyond the mean-field approximation, but it relies on the hypothesis that the full many-body state remains close to the initial state throughout the evolution -- in other words, the quantum fluctuations building up on top of the mean-field state are weak at all times. Since the initial state breaks the rotational symmetry of the system - and the evolved state inherits the symmetry breaking - this approach tacitly implies that symmetry breaking is observed at sufficiently low energy in the system, corresponding to the excitation energy of the mean-field ground state with respect to the actual ground state of the system. We shall explicitly test this assumption below. 
 
 Finally we observe that any finite spin system, although prepared in an initial state that breaks the symmetry of the Hamiltonian, is expected to undergo a depolarization dynamics which restores this symmetry, and this aspect is obviously not captured by our calculations. Nonetheless the explicit object of our study is not the evolution of the average collective spin - which is absent within our approach - but rather the evolution of the \emph{fluctuations} around the instantaneous average value, which is therefore tacitly assumed to be essentially independent of the average value; or alternatively it is assumed to occur on much faster timescales than the dynamics of the collective spin. This latter assumption is well justified in the limit $N\to\infty$, as the collective spin relaxes over time scales that grow linearly with system size -- reflecting a characteristic energy scale of order ${\cal O}(1/N)$ for the collective-spin excitations, related to the so-called Anderson's tower of states \cite{Anderson1952}. On the other hand, the frequencies of the LSW excitations do not scale with system size, so that the dynamics of fluctuations around the average value can completely decouple from that of the average in the $N\to\infty$ limit.
  
 \subsection{Details of the spin-wave approach} 
  
 To describe weak quantum fluctuations around the mean-field state one can introduce the linearized Holstein-Primakoff (HP) transformation, mapping spins onto bosons, and describing small deviations around the mean-field state: 
	\bearr
		S_i^x & = &  s - b_i^\dagger b_i \nonumber \\
		S_i^y & = & \sqrt{2s} ~ \frac{b_i - b_i^\dagger}{2i} \left[ 1+ {\cal O}\left( \frac{b_i^\dagger b_i}{2s} \right) \nonumber \right] \\
		S_i^z & = & -\sqrt{2s}~ \frac{b_i + b_i^\dagger}{2} \left [ 1+ {\cal O}\left( \frac{b_i^\dagger b_i}{2s} \right) \right ].
		\label{eq_HP_transform}
	\eearr 
	The linearization of the HP transformation to treat the dynamics of the system requires therefore that $\langle b_i^\dagger b_i \rangle (t) \ll 2s$ at all times. To study the dynamics we rewrite the Hamiltonian in terms of the HP bosons dropping all terms beyond quadratic in the $b^{(\dagger)}_i$'s operators, and obtaining (up to a constant energy that we neglect): 
	\be 
		{\cal H}^{(2)} = Js \sum_{i,j} (b_i^\dagger A_{ij} b_j + b_i A_{ij} b_j^\dagger + b_i B_{ij} b_j + b_i^\dagger B_{ij} b_j^\dagger)
 \ee	
 where 
 	\bearr
 		A_{ij} &=& \delta_{ij} \sum_{j'(\neq i)} r_{ij'}^{-\alpha} - r_{ij}^{-\alpha}/2 \nonumber \\
 		B_{ij} &=& r_{ij}^{-\alpha} / 2 ~.
 	\eearr
 	Assuming translational invariance we introduce the Fourier transformed Bose operators $b_{\bm k} = N^{-1/2} \sum_j e^{-i{\bm k}\cdot{\bm r}_j} b_j$, where $N$ is the number of spins, together with the Fourier transform of the interaction $\gamma_{\bm k} = \sum_{j} e^{i{\bm k}\cdot{\bm r}_j} r_{0j}^{-\alpha}$. With these notations, the Hamiltonian in momentum space takes the form: 
 	\be
 		 {\cal H}^{(2)} = (1/2) \sum_{k} \begin{pmatrix} b_{\bm k}^\dagger & b_{-\bm k} \end{pmatrix} 
 		 \begin{pmatrix}
 		 	A_{\bm k} & B_{\bm k} \\ B_{\bm k} & A_{\bm k}
 		 \end{pmatrix}
 		 \begin{pmatrix} b_{\bm k} \\ b_{-\bm k}^\dagger \end{pmatrix} 
 	\ee	
 	with $A_{\bm k} = 2sJ (\gamma_0 - \gamma_{\bm k}/2)$ and $B_{\bm k}= sJ \gamma_{\bm k}$.  The above Hamiltonian is then diagonalized by a Bogoliubov transformation \cite{blaizot-ripka} : $b_{\bm k}  = u_{\bm k} \beta_{\bm k} - v_{\bm k} \beta_{-\bm k}^\dagger$, with Bogoliubov coefficients $u_{\bm k} = {\rm sign}(b_{\bm k})[(A_{\bm k}/\omega_{\bm k} + 1)/2]^{1/2}$ and $v_{\bm k} = \sqrt{u_{\bm k}^2-1}$,  where we have introduced $\omega_{\bm k}= \sqrt{A_{\bm k}^2 - B_{\bm k}^2} = 2sJ \gamma_0 \sqrt{1 - \frac{\gamma_{\bm k}}{\gamma_0}}$. Since $\omega_{k=0}=0$, the symmetry-restoring $k=0$ mode cannot be treated on the same footing as the others. We have simply discarded the $k=0$ sector in all our calculations, a procedure which does not affect the results in the thermodynamic limit (an alternative procedure would correspond to gap the $k=0$ mode with a term which vanishes in the thermodynamic limit, providing similar results). Finally, we obtain an effective Hamiltonian (dropping a constant ground-state energy):
	\be
	 {\cal H}_{\rm eff} = \sum_{\bm k} \omega_{\bm k} \beta_{\bm k}^\dagger \beta_{\bm k} ~.
	\label{H_eff}
	\ee
	To obtain the physical properties at a given time $t$, we work in Heisenberg representation: 
	\bearr
		b_{\bm k}(t) &=& u_{\bm k} \beta_{\bm k}(t) - v_{\bm k} \beta_{-\bm k}^\dagger(t) \nonumber \\
		 &=& u_{\bm k} \beta_{\bm k}(0)e^{-i\omega_{\bm k} t} - v_{\bm k} \beta_{-\bm k}^\dagger(0) e^{i\omega_{\bm k} t} \nonumber \\
		 &=&  \left [ \cos(\omega_{\bm k} t)- i\frac{A_{\bm k}}{\omega_{\bm k}} \sin(\omega_{\bm k} t) \right ] b_{\bm k}(0) \nonumber \\
		 & &- i \frac{B_{\bm k}}{\omega_{\bm k}} \sin(\omega_{\bm k} t) b_{-\bm k}^\dagger(0) 
	\eearr	
	where on the last line we used $u_{\bm k}^2 = (1/2)(A_{\bm k} / \omega_{\bm k} + 1)$, $v_{\bm k}^2 = (1/2)(A_{\bm k} / \omega_{\bm k} - 1)$ and $2u_{\bm k} v_{\bm k} = B_{\bm k} / \omega_{\bm k}$. Since the initial state is the vacuum of the $b$'s operators, it trivially satisfies Wick's theorem \cite{blaizot-ripka} for the $b^{(\dagger)}(0)$ operators. As $b^{(\dagger)}(t)$'s operators are linear combinations of the $b^{(\dagger)}(0)$'s, Wick's theorem is then satisfied at any time. As a consequence, in a translationally invariant system all the information required to reconstruct the physical properties is contained in the one-body correlators: 
\bearr
	 \langle b_{\bm k}^\dagger b_{\bm k} \rangle(t) &=& \left ( \frac{B_{\bm k}}{\omega_{\bm k}}\right)^2 \sin^2 (\omega_{\bm k} t) \\
	 \langle b_{\bm k} b_{-\bm k} \rangle(t) 	 &=& -\frac{A_{\bm k} B_{\bm k}}{\omega_{\bm k}^2} \sin^2(\omega_{\bm k} t) - i \frac{B_{\bm k}}{2\omega_{\bm k}} \sin(2\omega_{\bm k} t)~.~~~~
	 \label{eq_correl_b}
	\eearr

\begin{figure}
	\includegraphics[width=\linewidth]{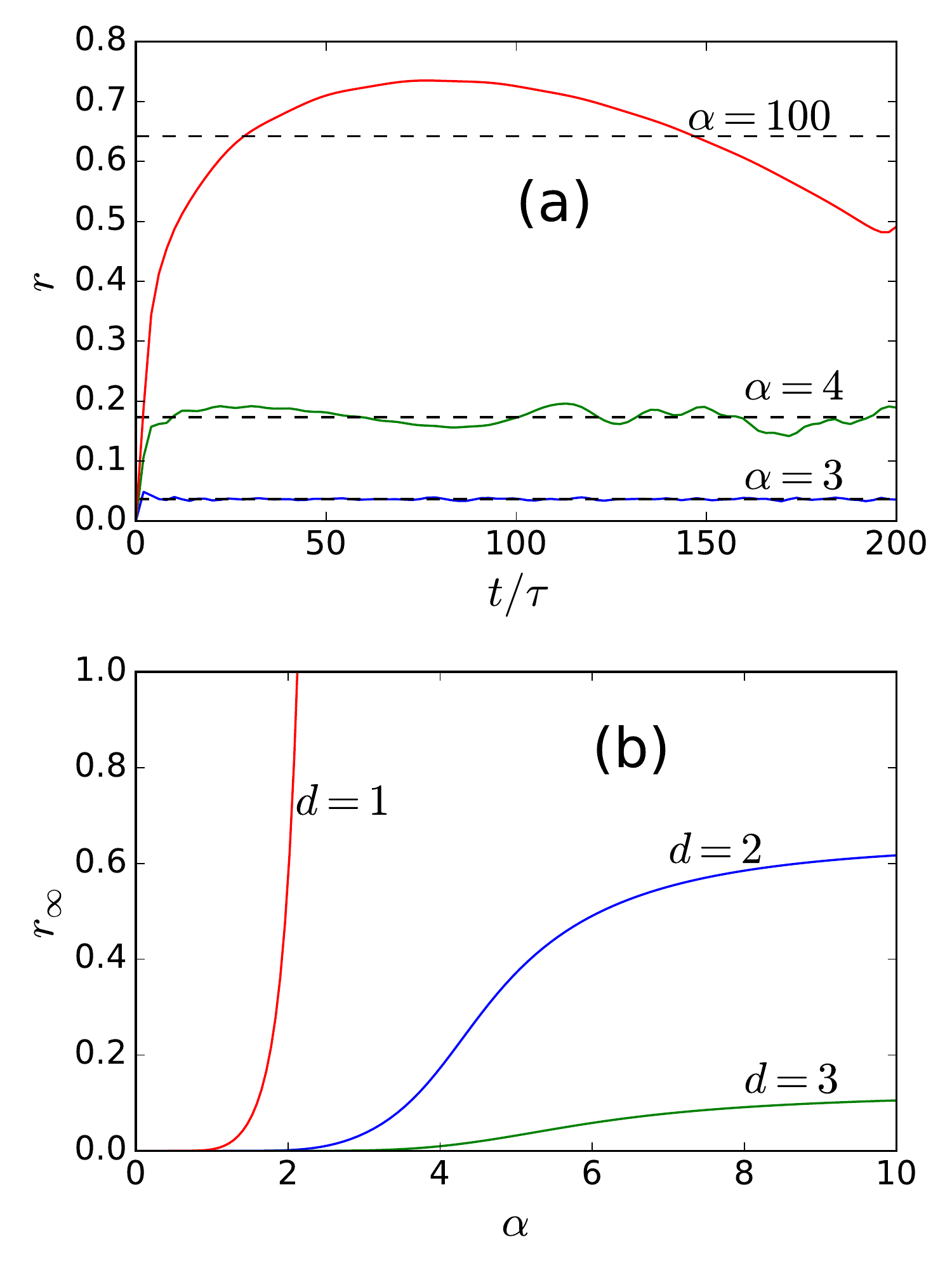}
	\caption{Stability of the spin-wave approximation ($s=1/2$). (a) $r$ of Eq. \eqref{eq_r} as a function of $t/\tau$ with $\tau=(2sJ\gamma_0)^{-1}$ in $d=2$ ($Lx=Ly=200$). Dotted lines are the value $r_\infty$ of Eq. \eqref{eq_r_inf}. (b) $r_\infty$, evaluated in $1d$ ($L=10000$), $2d$ ($Lx=Ly=200$) and $3d$ ($Lx=Ly=Lz=30$).}
	\label{fig_stability}
\end{figure}
	
\subsection{Maximum group velocity for $d+2 \lesssim \alpha $} 	
	
Here we calculate explicitly the divergence of the group velocity around $k=0$ when $\alpha$ approaches $d+2$ from above.
\begin{align}
& \frac{c(\alpha)}{2Js} = \lim_{\delta k \to 0} \frac{1}{\delta k} \sqrt{\gamma_0} \sqrt{\gamma_0 - \gamma_{(\delta k,0,0..)}} \nonumber \\
& = \lim_{\delta k \to 0} \frac{1}{\delta k} \sqrt{\gamma_0} \left ( \frac{(\delta k)^2}{2} \sum_{\bm r} \frac{x^2}{r^{\alpha}} +  ... \right)^{1/2} 
\sim \sqrt{\frac{1}{\alpha - d - 2}}
\end{align}	
where $kx= {\bm k} \cdot {\bm r}$, and the lattice sum has been approximated with an integral. The prefactor of the $1/\sqrt{\alpha -d-2}$ divergence depends on the dimension $d$. 
The above calculation shows the expected divergence for $\alpha \to (d+2)^+$. 
	
\subsection{Validity condition of spin-wave theory}

The LSW approach to the dynamics is justified under the condition
\be 
	r(t) = \frac{1}{2sN}\sum_{\bm k} \langle b_{\bm k}^\dagger b_{\bm k} \rangle (t) \ll 1 ~.
	\label{eq_r}
\ee
Since $\langle	b_{\bm k}^\dagger b_{\bm k} \rangle(t)$ does not depend on $s$, one immediately sees that the spin-wave approximation becomes exact in the limit of large $s$. 

The factor $r(t)$ is generally found to grow with $t$ (see Fig.~\ref{fig_stability}(a)), so that the worst-case scenario to test the consistency of LSW theory can be found in the limit $t \to \infty$. In this limit, as well as in the $N \to \infty$ one,  the $r$ factor takes the form:
	\bearr
		r(\infty) &\approx &\frac{1}{4sN} \sum_{\bm k} \left( \frac{B_{\bm k}}{\omega_{\bm k}} \right)^2 \label{eq_r_inf} \\
		&\approx & \frac{1}{4s}  \int_{\rm BZ} \frac{d^d k}{(2\pi)^d} \left( \frac{B_{\bm k}}{\omega_{\bm k}} \right)^2~,
	\label{stability_HP}
	\eearr
	where we have replaced $\sin^2( \omega_{\bm k} t)$ by $1/2$, since oscillations at nearby frequencies are dephased at long times and only the average survives. The momentum integration runs over the Brillouin zone (BZ). 

   Fig.~\ref{fig_stability}(b) shows $r(\infty)$ for finite-size lattices as a function of $\alpha$ in dimensions $d=1,2,3$, and in the extreme quantum case $s=1/2$. To put these results in perspective one can notice that: 
      \begin{enumerate}
      \item for $\alpha < d$, $B_{\bm k} / \omega_{\bm k} = (\gamma_{\bm k} / 2\gamma_0) / \sqrt{1 - \gamma_{\bm k}/\gamma_0}$ goes to zero when $N \to \infty$ at any nonzero $k$, since $\gamma_0 \to \infty$ while $\gamma_{\bm k}$ is finite.
On the other hand, $B_{\bm k} / \omega_{\bm k}$ has a finite limit for $k=0$. As a consequence $r(\infty)$ vanishes for $\alpha <d$, corresponding to a mean-field regime, identical (in the thermodynamic limit) to the $\alpha=0$ limit, where the dynamics occurs only in the collective spin $k=0$ sector;
\item for $\alpha > d$, if the exact value of the integral in Eq.~\eqref{stability_HP} must be evaluated numerically, one can easily identify the condition under which it is convergent in the infrared, validating the LSW approach. Since $B_{k=0}$ is finite, the condition requires $1/\omega_{\bm k}^2$ to grow at small $k$ more slowly than $k^{-d}$. If $\omega_{\bm k} \sim k^z$ with $z$ the dynamical exponent, the convergence requires $z < d/2$. As discussed in the main text, $z = 1$ for $\alpha \ge d+ 2$ and $z = (\alpha - d) / 2$ for $d \le \alpha \le d+2$ \cite{paper1-XX_LR}. The convergence is hence always guaranteed in $d=3$, in $d=2$ for $\alpha < 4$ (with a logarithmic divergence for $\alpha \ge 4$), and in $d=1$ for $\alpha < 2$. Due to the weak logarithmic divergence of $r(\infty)$, LSW remains quantitative on sufficiently small sizes for $\alpha=2$ in $d=1$, and for $\alpha>4$ in $d=2$.
\end{enumerate}
We notice that the above conditions of validity of LSW theory in the pre-thermalized state are more restrictive than the condition that the theory has to satisfy at zero temperature -  which reads $z < d$, always satisfied in $d=2$ and 3, and satisfied for $\alpha < 3$ in $d=1$. On the other hand, the condition $z<d/2$ is the same as the one guaranteeing the validity of LSW theory at \emph{finite} temperature $T$ -- which requires the thermal correction to the order parameter
\begin{equation}
\frac{1}{N} \sum_{\bm k} (u_{\bm k}^2 + v_{\bm k}^2) \langle \beta_{\bm k}^{\dagger} \beta_{\bm k} \rangle \approx \frac{k_B T}{N}  \sum_{\bm k} \frac{A_{\bm k}}{\omega_{\bm k}^2}
 \end{equation}
to be finite. Even though the pre-thermalized state is not a thermal one, it shares with a thermal state the property of being at finite energy density. 
      

\subsection{Correlation functions and entanglement in LSW theory}
Here we provide more details on the LSW expression for the various correlation functions studied in this work. Making use of Eqs. \eqref{eq_correl_b} and \eqref{eq_HP_transform}, the structure factor for the $z$ component of the magnetization takes the form:
\bearr 
	S_k^{zz} &=& \frac{1}{N} \sum_{ij} \langle S_i^z S_j^z \rangle e^{ik(r_j - r_i)} \nonumber \\
	 &\approx& \frac{s}{2} \langle (b_{\bm k} + b_{-\bm k}^\dagger)(b_{-\bm k} + b_{\bm k}^\dagger) \rangle \nonumber  \\
	 &=& \frac{s}{2} \left[ 2 \Re \langle b_{\bm k} b_{-\bm k} \rangle + 2 \langle b_{\bm k}^\dagger b_{\bm k} \rangle  + 1\right] \nonumber  \\
	 &=& \frac{s}{2} \left[ 1 - \frac{\gamma_{\bm k}}{\gamma_0} \sin^2(\omega_{\bm k} t)  \right]
	 \label{eq_Sk_zz_tot}
\eearr
	 which corresponds the expression given in Eq.~3 of the main text. A similar calculation for the $yy$ correlations gives: 

\be 
	S_k^{yy}(t) = \frac{s}{2} \left[ 1 + \frac{\gamma_{\bm k}}{\gamma_0 - \gamma_{\bm k}} \sin^2(\omega_{\bm k} t)  \right]
	 \label{eq_Sk_yy_tot}
\ee

The calculation of the entanglement entropy within LSW theory is well established, and based on the evaluation of the correlation matrix -- we refer the reader to Ref.~\cite{FrerotR2015} for the details.

\section{Quantum field-theory description of the long-range XX model}

The XX model is a fundamental model of strongly interacting bosons with translational as well as particle-hole symmetry. As it displays a gapless spectrum and algebraic correlations, its long-wavelength properties are expected to be correctly described via a quantum field theory in continuum space. In the case of short-range interactions the quantum-field theory corresponds to the quantum O(2) model for a complex scalar field $\psi({\bm r},\tau)$, whose action reads
 \begin{equation} 
 {\cal A}  = \int_0^{\beta} d\tau \int d^d x \left (  |\partial_\tau \psi|^2 + |{\bm \nabla} \psi|^2 + r|\psi|^2 + u|\psi|^4 \right )~.
 \label{e.action}
 \end{equation} 
 The above action can be obtained explicitly \cite{Fisheretal1989, Sachdev2011} by mapping the quantum spins onto bosons via an HP transformation (with quantization axis along $z$) giving rise to a Bose-Hubbard-like model; using the coherent-state path-integral representation for the partition function of the bosonic model; and applying a Hubbard-Stratonovich transformation via the introduction of the auxiliary field $\psi$.    
 The case of long-range interactions decaying as $1/r^\alpha$ in the spin model corresponds to long-range hopping in the bosonic model. The Hubbard-Stratonovich transformation leads to an effective action for the auxiliary field with a quadratic part given by
 \cite{Sachdev2011} 
 \begin{equation}
 \int d\tau ~ \sum_{ij} \psi^*_i(\tau) ~\tilde{J}^{-1}_{ij} ~\psi_j(\tau) 
\end{equation}
 where 
 \begin{equation}
 \tilde{J}^{-1}_{ij} = \frac{1}{N} \sum_{\bm k} \frac{e^{i{\bm k}\cdot({\bm r}_i - {\bm r}_j)}}{K + \gamma_{\bm k}} 
 \end{equation}
 where $K$ is a positive constant such that $K + \gamma_{\bm k} > 0$, namely $K > \max_{\bm k} |\gamma_{\bm k}|$, and $\gamma_{\bm k}$ is the Fourier transform of the $1/r^{\alpha}$ interaction. A Taylor expansion of the numerator gives then
  \begin{align}
 & \int d\tau ~\sum_{ij} \psi^*_i(\tau) ~\tilde{J}^{-1}_{ij} ~\psi_j(\tau)  \nonumber \\
 & =  \frac{1}{K} \int d\tau \left [ \sum_i  |\psi_i|^2 - \frac{1}{K} \sum_{ij}  \frac{\psi^*_i \psi_j}{r_{ij}^{\alpha}} + {\cal O}(\gamma_{\bm k}/K)^2 \right ] \nonumber \\
 & \approx \int d\tau \int d^d {\bm r} \left  ( a  |\psi|^2 + b  |{\bm \nabla} \psi|^2 + ...  \right ) \nonumber \\ 
 & - v \int d\tau \int d^d r~ d^d r' ~ \frac{\psi^*(\bm r,\tau) \psi (\bm r',\tau)}{|\bm r - \bm r'|^{\alpha}}
\end{align}
In the last term the first part (containing a single spatial integral) reproduces the short-range part of the coupling for the auxiliary field, while the double integral captures the long-range part. 
Therefore, upon proper rescaling of the coefficients, the inclusion of long-range interactions amounts to adding the $v$ term to the effective action of Eq.~\eqref{e.action}. 

Moving to $k$-space and Matsubara frequencies, the action of the long-range interacting system takes the form \cite{DuttaB2001}
\begin{eqnarray}
{\cal A} &=& \int d\omega \int \frac{d^d k}{(2\pi)^d} \left [ q^2 + \omega^2 + r + v' q^\sigma \right ] |\psi(\bm k, \omega)|^2 \nonumber \\
&+& {\cal A}_{\rm int}
\end{eqnarray}
where ${\cal A}_{\rm int}$ contains the terms beyond quadratic, and $\sigma = \alpha -d$. Far from its quantum critical point, the action can be treated within the Gaussian approximation \cite{Kardar-book} (which represents the field-theory analog to the LSW 
approach we relied upon); in this case a RG treatment by simple dimensional analysis shows that the long-range interaction term $v' q^\sigma$ is relevant (and the $q^2$ term irrelevant) when $\sigma<2$, namely for $\alpha < d+2$. Under this circumstance the quantum field theory loses therefore Lorentz invariance, and $z = \sigma/2<1$ acquires the role of new dynamical critical exponent.    

As it is not our purpose to investigate the quantum critical point of the long-range quantum O(2) model, we can stop our analysis at the level of the Gaussian approximation. In the case of the long-range quantum O(1) model  -- describing the quantum phase transition of the long-range Ising model in a transverse field -- a perturbative RG analysis can be found in Ref.~\cite{maghrebietal2016}.

\section{Pre-thermalization dynamics in a subsystem}
 The LSW approach, neglecting interactions among the spin waves, is unable to describe a true thermalization process. 
 This is explicitly manifested in the expressions in Eq. \eqref{eq_correl_b}, \eqref{eq_Sk_zz_tot} and \eqref{eq_Sk_yy_tot}: the different Fourier components of the structure factor show persistent oscillations at a frequency $2\omega_{\bm k}$ without any form of relaxation. Nonetheless, if one looks at the structure factor \textit{in a subsystem}, the dephasing of oscillations at nearby frequencies leads to a \textit{pre-}thermalization process, driving the subsystem towards a generalized Gibbs ensemble \cite{VidmarR2016}, with a different (generalized) ``temperature'' for each $k$ component. It is the purpose of this section to provide additional details on this pre-thermalization dynamics. 

\begin{figure}
\includegraphics[width=\linewidth]{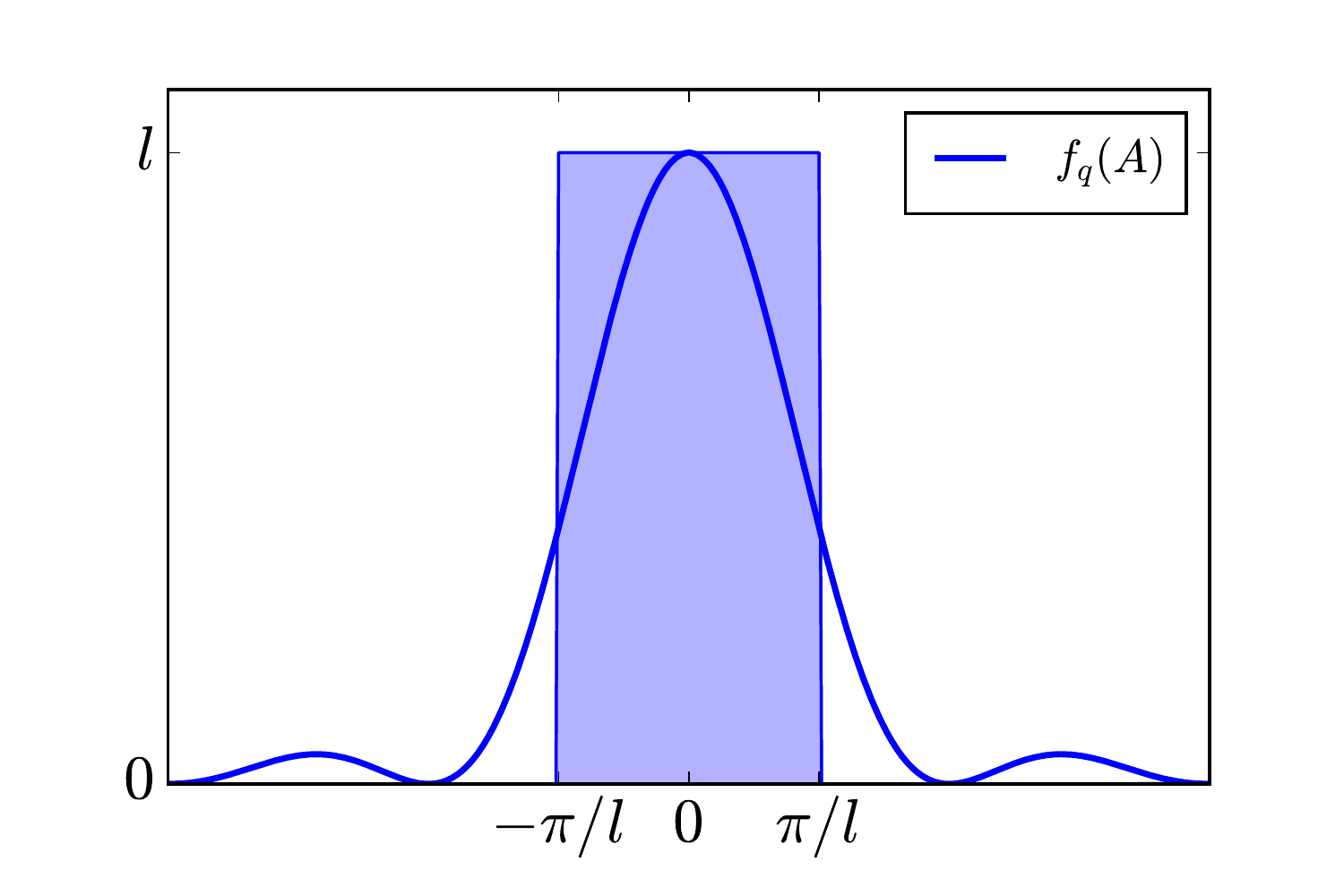}
\caption{Form factor $f_A(q)$ (Eq. \eqref{eq_fq}) in $d=1$, here for $l_A=100$ and $L=10^4$. When $A$ is large, one can approximate $f_A(\bm q)$ by a step-like function enclosing the shaded area, which has the same integral $2\pi/l_A$ as the $f_A$ function.}
\label{fig_fq}
\end{figure}

\subsection{Relaxation of the fluctuations in a subsystem}
\label{sec_relaxation_fluctuations_subsystem} 
Considering a subsystem $A$, it is convenient to introduce a \textit{weight function} $w_A(i)$, weighting sites $i$ belonging to $A$. If $A$ contains $N_A$ sites, the most obvious choice is $w_A(i) = 1/\sqrt{N_A}$ if $i \in A$ and $0$ otherwise, but the following expressions are more general. The structure factor for the fluctuations of the spin component $S^{\beta}$ in the subsystem $A$ then reads : 
\bearr
	S^{\beta\beta}_A(\bm k) &=& \sum_{ij} w_A(i) w_A(j)  \langle S^{\beta}_i S^{\beta}_j \rangle e^{i\bm k\cdot(\bm r_j - \bm r_i)}  \nonumber \\
	&=& \frac{1}{N} \sum_{ij} w_A(i) w_A(j)  \sum_{q} S^{\beta\beta}(\bm q) e^{i(\bm k-\bm q)\cdot(\bm r_j - \bm r_i)} \nonumber \\
	&=& \frac{1}{N} \sum_{q} f_A(\bm q) S^{\beta\beta}(\bm k + \bm q)
	\label{eq_SkA}
\eearr
where we have introduced the form factor
\be 
	f_A(\bm q) = \left |  \sum_j e^{i\bm q \cdot \bm r_j } w_A(j) \right | ^2 ~.
\ee
Typically, if $A$ has linear size $l_A$, $f_A(\bm q)$ selects wavectors in the range $k \pm \Delta k$ with $\Delta k \approx \pi / l_A$. 

Let us first consider the case $d=1$, postponing the case $d>1$ to the next subsection. If $w_A(i)$ selects sites inside a segment of length $l_A$, one has 
\be 
	f_A(q) = \frac{1}{l_A} \frac{\sin^2(ql_A/2)}{\sin^2(q/2)}
	\label{eq_fq}
\ee
and $f_A(k=0) = l_A$. The relaxation of the fluctuations at wavevector $k$ in a subsystem then results from the dephasing of nearby frequencies around the ``carrier" $\omega_k$. One expects such a dephasing mechanism to be effective after a timescale given by the inverse bandwidth of the frequencies involved in the subsystem dynamics, namely
\bearr
	t^*_k  \sim  \frac{1}{2\vert \omega_{k+\Delta k} - \omega_{k} \vert } 
	\sim  \frac{1}{2\Delta k \vert \partial \omega_{k} / \partial k \vert}
	 \sim  \frac{l_A}{2\vert v_G(k) \vert}
\eearr
with $v_G(k) = d \omega_k / d k$ the group velocity. One thus recovers the typical timescale for the damping of subsystem fluctuations at wavevector $k$ as discussed in the main text. 

More quantitatively, in the limit $N \to \infty$, one has 
\be 
S^{\beta\beta}_A(k) = \int_{\rm BZ} \frac{dq}{2\pi} f_A(q) S^{\beta\beta}(k + q) 
\ee 
For a sufficiently large $A$ subsystem, $f_A(q)$ is peaked around $q=0$, with $\int_{\rm BZ} (dq/2\pi) f_A(q) = 1$. It is therefore legitimate to approximate $f_A(q)$ by a simple step function $f_A(q) = l_A$ for $-\pi / l_A < q < \pi / l_A$ and $f_A(q)=0$ otherwise, as illustrated on Fig. \ref{fig_fq}.

Within this approximation, one obtains
\be 
	S^{\beta\beta}_A(k) = \frac{l_A}{2\pi}\int_{-\pi/l_A}^{\pi/l_A} dq ~S^{\beta\beta}(k + q)  ~.
	\label{e.Sbetabeta}
\ee
From Eqs. \eqref{eq_Sk_zz_tot} and \eqref{eq_Sk_yy_tot}, one sees that the structure factor $S^{\beta\beta}(k)$ has the form 
\be 
	S^{\beta\beta}(k,t) = \bar{S}^{\beta\beta}(k) + g^{\beta\beta}_k \cos(2 \omega_{k} t)~.
\ee
E.g. for $\beta=z$ we obtain: 
\bearr
	\bar{S}^{zz}(k) &=& \frac{s}{2} \left(1 - \frac{\gamma_{k}}{2 \gamma_0} \right) \nonumber \\
	g_k^{zz} &=& \frac{s\gamma_{k}}{4\gamma_0} ~.
\eearr
As the strongest dependence of $S^{\beta\beta}(k+q,t)$ on $q$ comes from the cosine term, in the integral over $q$ we can approximate all other $q-$dependent terms with their value at $q=0$. This reduces the calculation of Eq.~\eqref{e.Sbetabeta} to the following integral: 
\begin{align}	
	& \frac{l_A}{2\pi}\int_{-\frac{\pi}{l_A}}^{\frac{\pi}{l_A}} dq \cos( 2 \omega_{k+q} t) \approx   \Re \left[ \frac{l_A}{2\pi} \int_{-\frac{\pi}{l_A}}^{\frac{\pi}{l_A}} dq ~e^{ 2it( \omega_k + q \partial_k \omega_k)} \right] \nonumber \\
	& = \cos(2 \omega_{k} t) ~\frac{\sin(2\pi \partial_k \omega_{k} t / l_A)}{2\pi \partial_k \omega_k t / l_A}~.
\end{align} 
This leads to Eq.~(5) of the main text (for the case $d=1$).
We observe that the damping term ${\rm sinc}(\pi t/t_k^*)$ comes from the dephasing of nearby frequencies, leading to an effective pre-thermalization within the subsystem. 

\begin{figure}
	\includegraphics[width=\linewidth]{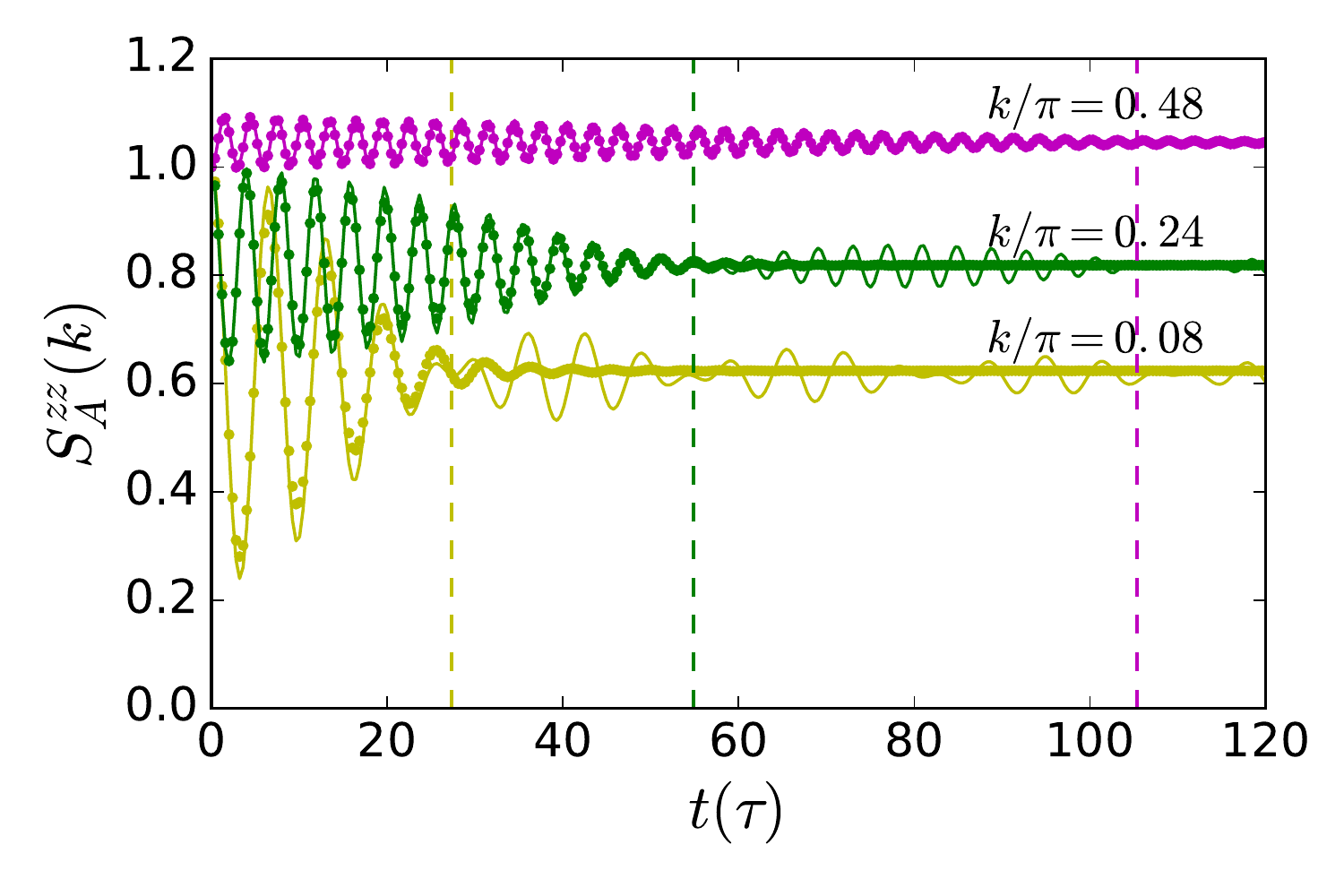}
	\caption{Relaxation of the subsystem structure factor in $d=1$. Here, $L=10^4$, $l_A=50$ and $\alpha=2$. Dots are the exact expression from Eq. \eqref{eq_SkA}, while lines are the approximate expression from Eq. 5 of the main text. Time is expressed in units of $\tau = (2sJ\gamma_0)^{-1}$. Vertical dashed lines mark the times $t^*_k = l / 2v_G(k)$.
	}
	\label{fig_Ansatz_SkA}
\end{figure}

As illustrated in Fig. \ref{fig_Ansatz_SkA}, Eq.~(5) of the main text reproduces perfectly the relaxation dynamics up to $t=t^*_k$. At $t>t^*_k$, the approximate expression of Eq.~(5) predicts small revivals of the oscillations which are not observed with the complete expression of Eq. \eqref{eq_SkA}. 

\subsection{Multi-speed relaxation for $d>1$}

The above considerations generalize readily to the case $d > 1$.  Considering a (hyper-)tetragonal subsystem of size $l_{A,1} \times l_{A,2} \times \dots$, one obtains Eq.~(5) of the main text, which we repeat here for completeness:
\be 
	S^{zz}_A(\bm k) = S_A^{zz}(\bm k, \infty) + g_{\bm k} \cos(2 \omega_{\bm k} t)\prod_{a=1}^{d}  {\rm sinc}[\pi t / t_{\bm k,a}^*]
	\label{eq_Ansatz_S_kA_higher_dims}
\ee
with $t_{\bm k,a}^* = l_{A,a} / (2 \partial_{k_a} \omega_{\bm k})$. The sinc term leads therefore to damping of the oscillations on a characteristic time scale $t^*_{\bm k} = \min_a t_{\bm k,a}^*$. An important qualitative difference with respect to the $1d$ case is that, even if the dispersion relation is linear at low frequency, $\omega_{\bm k} = c k$ (as it is found for $\alpha > d+2$), the group velocity has several components $v_{G,a} ({\bm k}) = \partial_{k_a} \omega_{\bm k} = c k_a / k$, and the minimal $t_{\bm k,a}$ depends on the orientation of the ${\bm k}$ vector with respect to the coordinate axes of the subsystem. In the case of an hypercube $l_A^d$, one has
\be 
	t^*_{\bm k} = \frac{l_A}{2c}~ \frac{k}{\max_a k_a}
\ee
which, depending on the orientation of ${\bm k}$, ranges between $l/2c$ and $(l/2c) \sqrt{d}$.

\subsection{Scaling forms for the evolution of the collective spin on a subsystem}

\begin{figure}
 	\includegraphics[width=\linewidth]{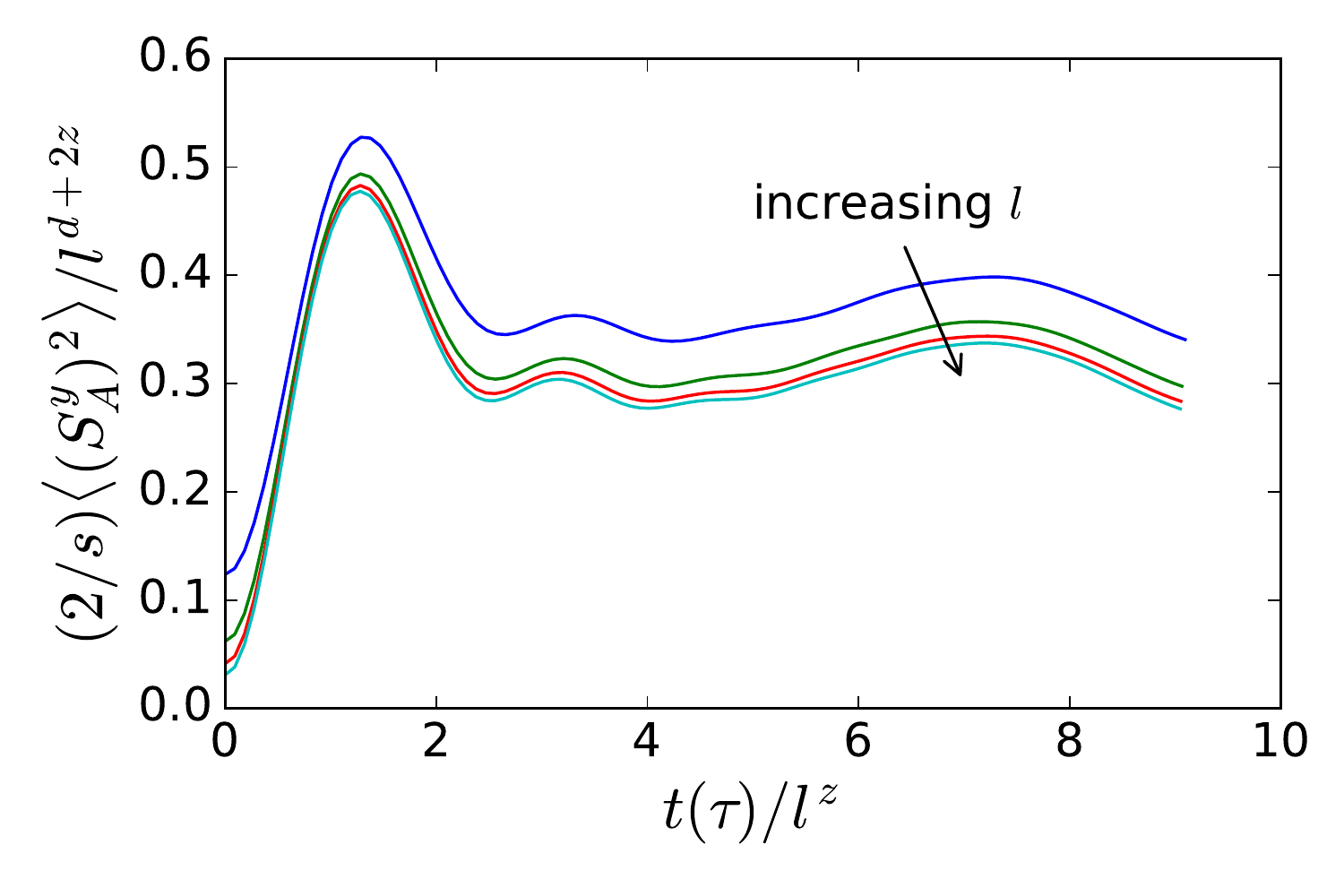}
 	\caption{Scaling of the variance of the collective spin $S_A^y$ in $d=2$ for $\alpha=3$. $A$ is a square region of size $l \times l$ in a torus of size $10l \times 10l$, with $l=8, 16, 24, 32$. Time (in units of $\tau = (2sJ\gamma_0)^{-1}$) is rescaled to $l^z$, while $\langle (S_A^y)^2 \rangle$ is recaled to $l^{d +2z}$, with $d=2$ and $z=1/2$. This plot should be compared with Fig. 3(b) of the main text where the variance of $S_A^z$ is plotted.
 	}
 	\label{fig_scaling_varSy_alpha_3}
\end{figure}

We have seen in the main text that the evolution of the subsystem structure factor at ${\bm k}=0$ for the $z$ spin component is compatible with the following Ansatz
\be 
     \langle (S_A^z)^2 \rangle = l_A^d~ f_z(t / l_A^z)
\ee
for the evolution of the variance of the collective spin component $S_A^z$ -- namely, a volume-law scaling at all times (including the initial state).  
 
 On the other hand, the variance of $S_A^y$ is found to relax towards a super-extensive value $\propto l_A^{d+2z}$ -- while it obeys a volume-law scaling in the initial state at $t=0$. At thermal equilibrium LSW predicts the same super-extensive scaling, which is therefore expected to hold for temperatures compatible with spontaneous symmetry breaking (namely, the hypothesis underlying LSW theory). The following scaling Ansatz 
\be 
	\langle (S_A^y)^2 \rangle(t) - \langle (S_A^y)^2 \rangle(0) = l_A^{d+2z} f_y(t/l_A^z)
\ee
regroups together this observation as well as the divergence of the group velocity, and it is found to be verified by the LSW data, as illustrated on Fig. \ref{fig_scaling_varSy_alpha_3} for $\alpha=3$ in $d=2$.
Incidentally, this scaling form demands that $2z$ be smaller than $d$, as it is impossible that the collective spin fluctuations scale faster than the square of the volume, which the same criterium of stability of LSW approximation, required by the convergency of the $r_\infty$ in Eq. \eqref{eq_r_inf}.

\begin{figure}
 	\includegraphics[width=\linewidth]{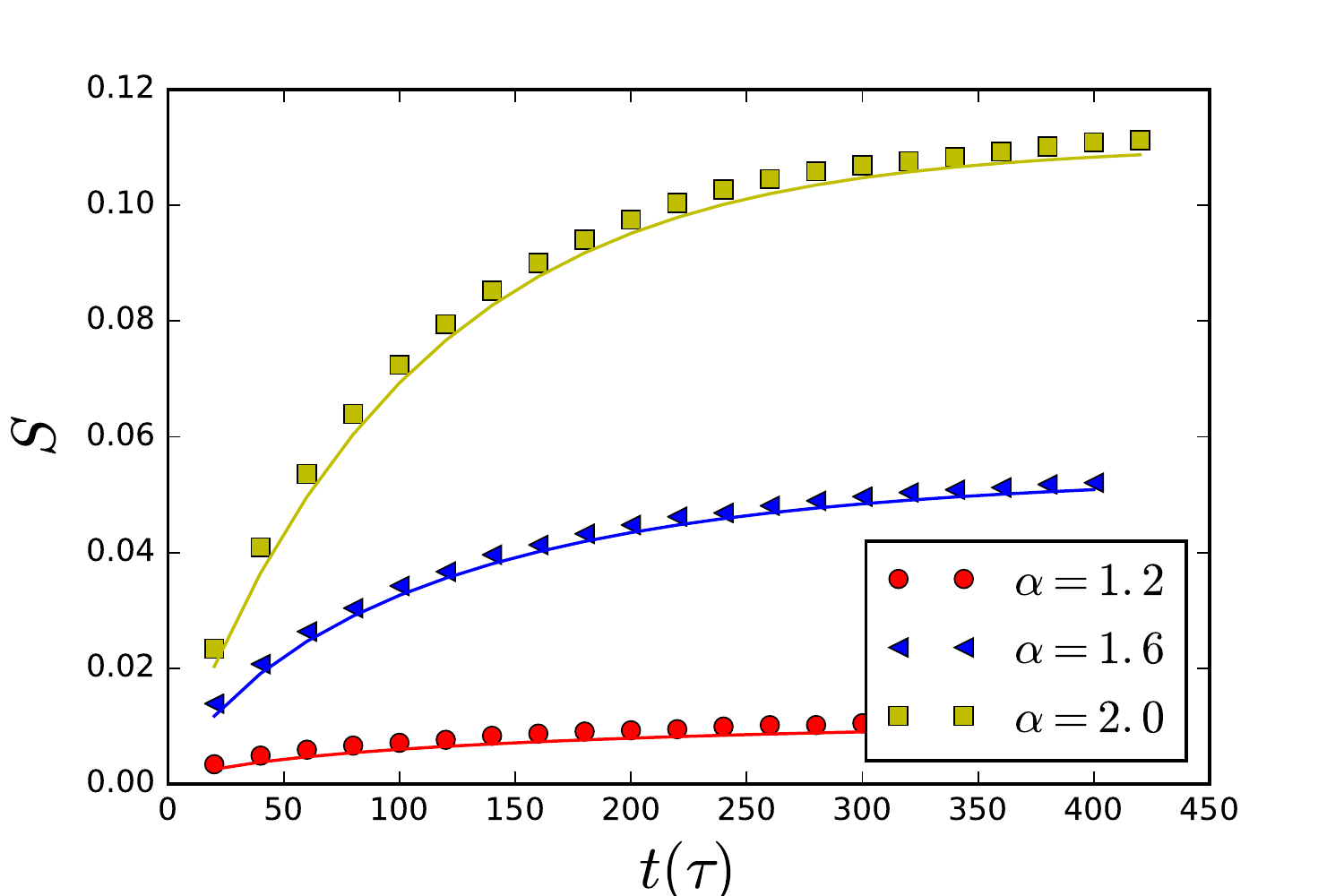}
 	\caption{Test of the mode-decomposition Ansatz for the growth of entanglement entropy in $d=1$. $A$ is a subsystem of length $l_A=400$ in a total system of size $L=16000$. Symbols represent the LSW results, while solid lines are the predictions of Eq. ~(6) of the main text.}
 	\label{fig_Ansatz_SA_1d}
\end{figure}

\begin{figure}
	\includegraphics[width=\linewidth]{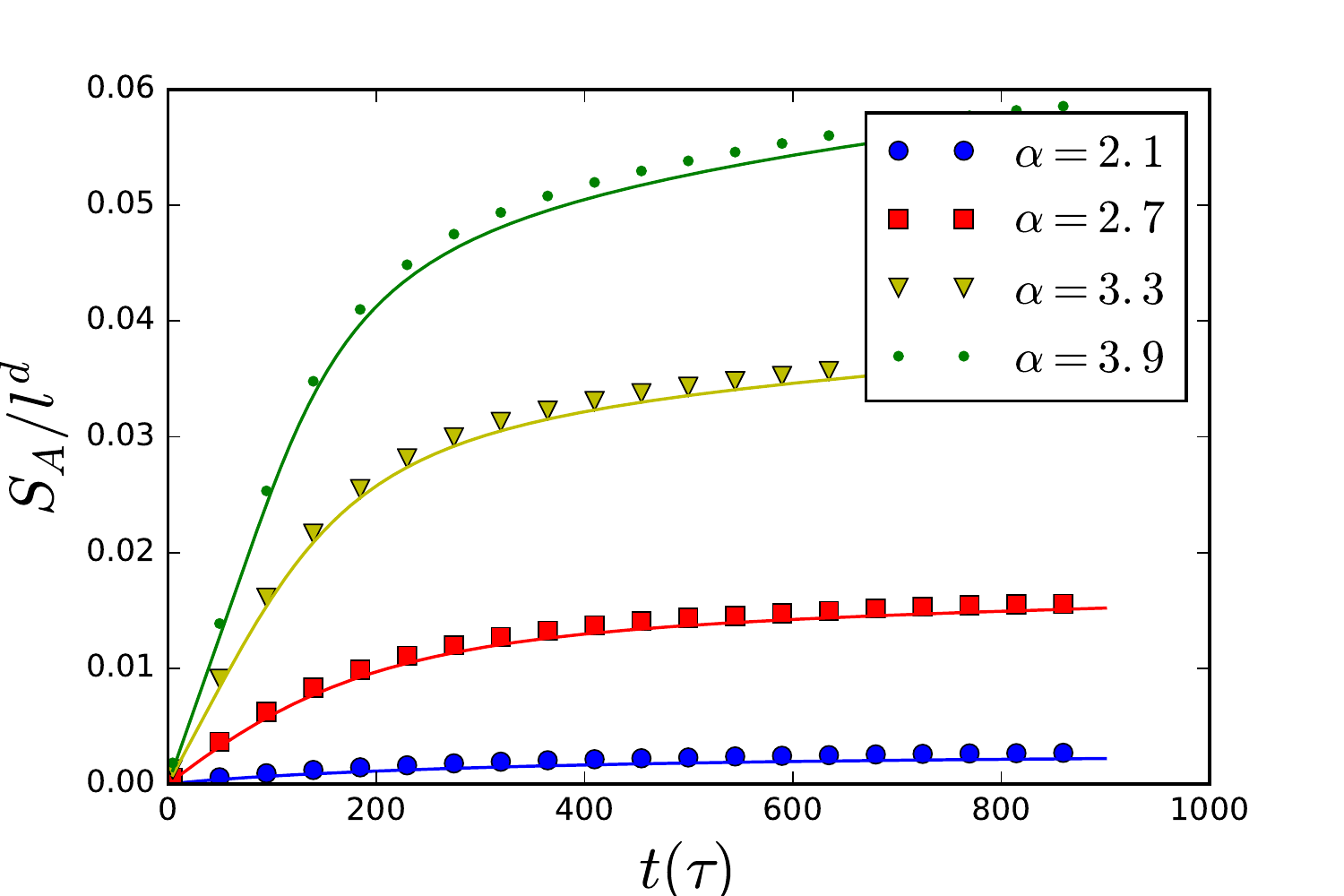}
 	\caption{Test of the mode-decomposition Ansatz for the growth of entanglement entropy in $d=2$. $A$ is a $150 \times 150$ cylinder in a $150 \times 1500$ torus. Lines and symbols as in Fig.~\ref{fig_Ansatz_SA_1d}.}
 	\label{fig_Ansatz_S_2D}
\end{figure}
\begin{figure}	
 	\includegraphics[width=\linewidth]{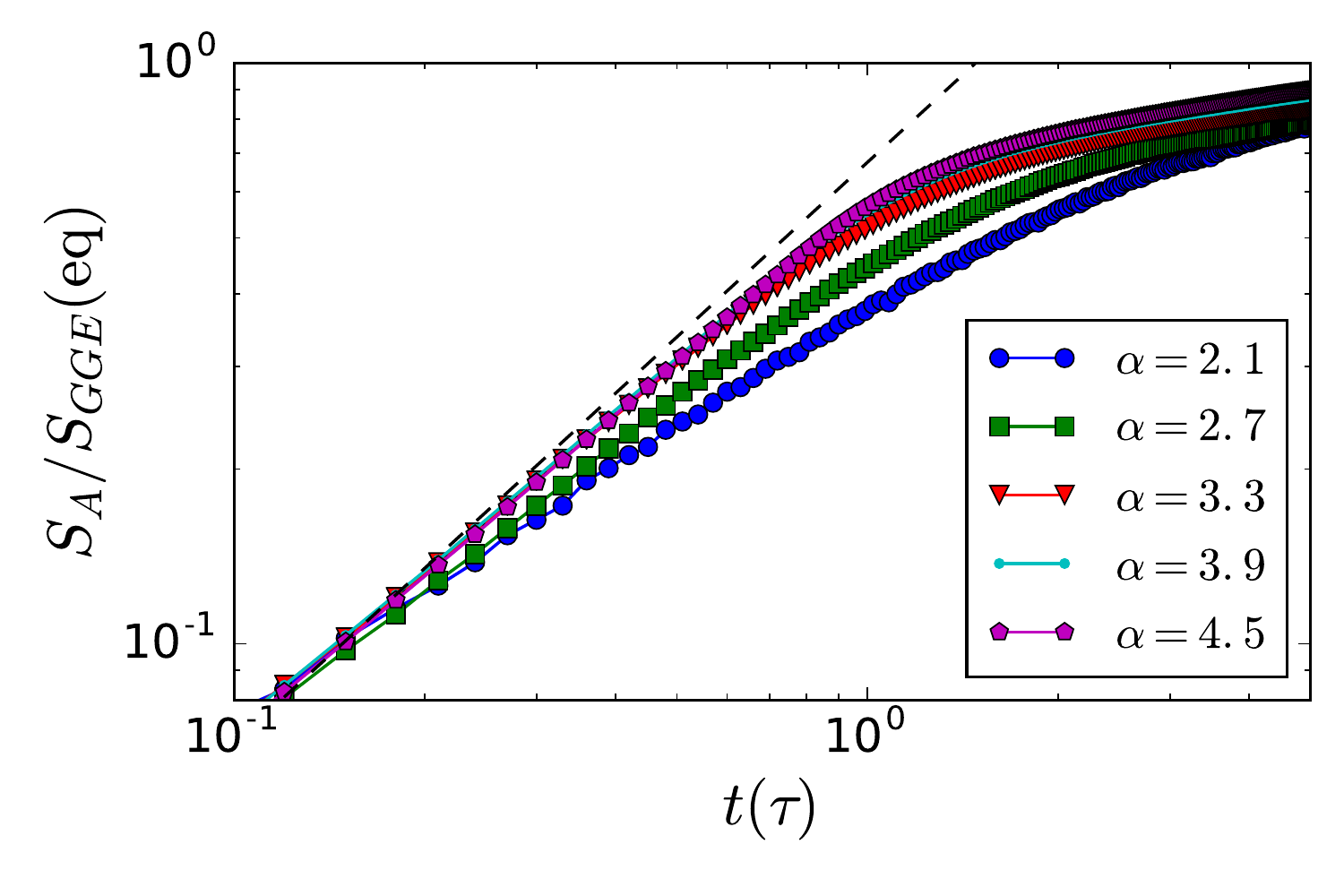}
 	\caption{Power-law growth of entanglement entropy in time for $d=2$. $A$ is a $150 \time 150$ cylinder in a $150 \times 1500$ torus. Entropy is rescaled to the asymptotic value expected from the GGE prediction (see text). Dashed line shows $S\sim t$.
 	}
 	\label{fig_scaling_S_2D}
\end{figure}

\subsection{Mode-decomposition Ansatz  and short-time behavior for the entanglement entropy}
\label{sec_relax_SA}

In this section we provide more details on the mode-decomposition Ansatz for the entanglement entropy, Eq.~(6) of the main text. Its strict validity - even in the context of the LSW approximation - can only be assumed in the limit in which subsystem $A$ becomes very large, so that the subsystem excitations can be identified with those excitations of the bulk system whose wavelength is commensurate with the linear dimensions of the subsystem itself; and the populations of the subsystem modes can be deduced from those of the bulk excitations, as fixed by the initial conditions of the quench. 
 
  In order to minimize the finite-size effects of subsystem $A$, its ideal geometry is that of a slice of a $L_1 \times L_2^{d-1}$ hyper-torus of size $l_A \times L_2^{d-1}$ ($l_A < L_1$). In this case finite-size limitations to the applicabiity of Eq.~(6) only appear in the "short" direction (of length $l_A$) which is perpendicular to the $A$-$B$ cut - this particular geometry is also very helpful in the numerical calculation of the entanglement entropy, as the correlation matrix necessary to reconstruct it \cite{FrerotR2015} is block-diagonal in the wavevector longitudinal to the cut.  In this geometry, for $d>1$ the time scale $t^*_{\bm k}$ associated to the buildup of the entropy in mode ${\bm k}$ is to be modified to $t^*_{\bm k,\perp} = l_A / 2|v_{\perp}|$, where $v_{\perp} = \partial_{k_\perp} \omega_{\bm k}$ is the component of the group velocity perpendicular to the $A$-$B$ cut; indeed only the propagation of the modes perpendicular to the cut is responsible for the entangling dynamics of the $A$ and $B$ subsystems. Figs.~\ref{fig_Ansatz_SA_1d} and \ref{fig_Ansatz_S_2D} confirm the validity of the decomposition Ansatz, as well as the choice of the scaling function $f$, for $d=1$ and $d=2$ respectively. 

The linear dependence of the function $f(t/t^*_{\bm k})$ on time may na\"ively suggest that the overall time dependence of the entanglement entropy at short times is always linear. This is generally true if the time scale $t^*_{\bm k,\perp}$ admits a non-zero lower bound $t^*_{\rm min}>0$: for $t<t^*_{\rm min}$ all modes contribute a linearly growing entropy up to time $t^*_{\rm min}$. 
Such a condition is verified when $z=1$, namely for $\alpha > d+2$: there $t^*_{\rm min}=l_A/(2c)$. In this case, Eq.~(6) of the main text shows indeed that the entropy grows linearly in time, with a prefactor proportional to the boundary of $A$, $S_A(t) \sim l_A^{d-1} t$.
On the other hand when $\alpha < d+2$ $c(\alpha) \to \infty$ so that the lower bound on the time scales vanishes, $t^*_{\rm min} \to 0$. This means that for arbitrarily short times (in the thermodynamic limit) one can always find some modes ${\bm k}$ that contribute non-linearly to the entanglement entropy, as their contribution has already saturated (for $t > t_{\bm k, \perp}$). Therefore in this regime one can expect a \emph{sublinear} growth of entanglement. 

The above expectations appear to be verified in Fig.~\ref{fig_scaling_S_2D}. There we observe that the short-time evolution of the entanglement entropy in a $d=2$ system appears generally compatible with a power-law behavior $S_A \sim t^x$.
If $\alpha > 4$, a linear growth of entanglement entropy is clearly observed at short times; in particular, when rescaling time to $\tau = l_A/(2Js\gamma_0)$ -- where $2Js\gamma_0$ sets the bandwidth of the dispersion relation -- the short time behavior becomes nearly independent of $\alpha$. On the other hand for $\alpha < 4$ our data clearly indicates that $x<1$.

 \begin{figure}[t]
 \includegraphics*[width=\linewidth]{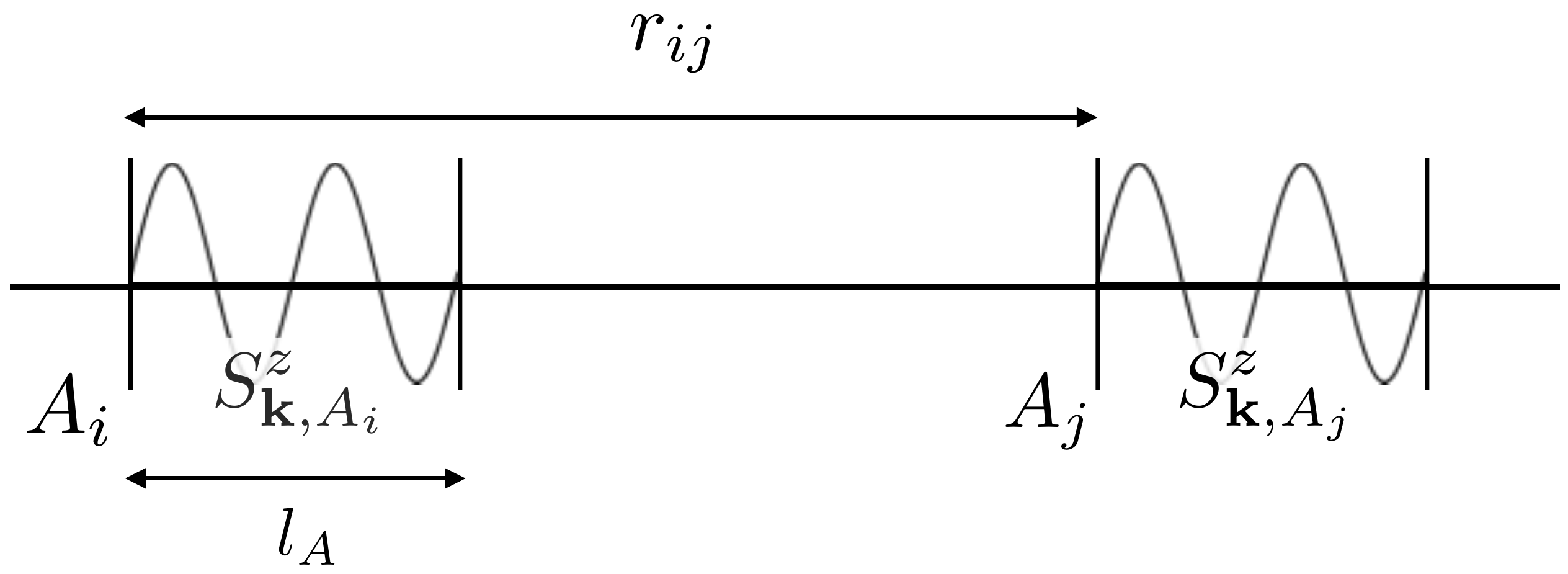}
  \caption{Sketch of the $C^{zz}_{\bm k}(A_i, A_j)$ correlations between two subsystems.}
 \label{f.Ckr}
\end{figure}	
 
 	\begin{figure*}[t]
 \includegraphics*[width=\linewidth]{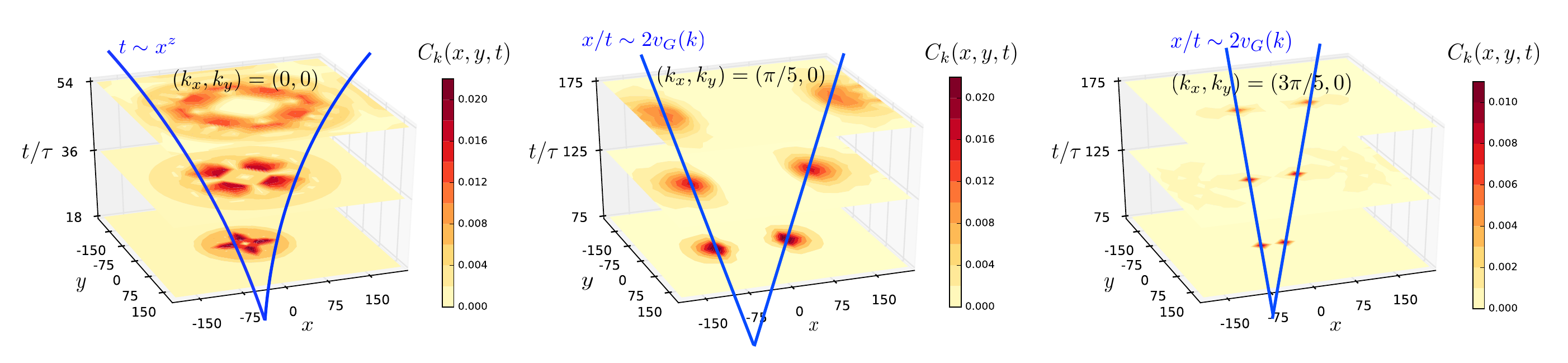}
  \caption{Space-time visualization of the multi-speed correlations spreading with $\alpha=3$ interactions in $d=2$. False colors represent correlations $C^{zz}_{\bm k}(\bm r,t)$  for $A_1$ and $A_2$ corresponding to two $10 \times 10$ subsystems (within a $420 \times 420$ total system) connected by the distance vector $\bm r=(x,y)$. Time is expressed in units of $\tau=(2Js\gamma_0)^{-1}$. (a) The $\bm k=(0,0)$ correlations spread in time with a superballistic front propagation, $r \sim t^2$; they develop a nearly spherical front at long times. (b) Correlations at $\bm k=(\pi/5,0)$ and (c) $\bm k=(3\pi/5, 0)$. At finite $k$, we observe a ballistic spreading $r \sim  2v_G(\bm k)t$ along the direction of $\bm v_G(\bm k)$.
  }
 \label{fig_Ck}
\end{figure*}

\section{Buildup of correlations among distant extended subsystems} 

 In this section, we provide additional details concerning the space and time dependence of the ${\bm k}$-dependent correlation function $C^{zz}_{\bm k}(A_i, A_j) = \langle \delta S^z_{\bm k,A_i} \delta S^z_{-\bm k,A_j} \rangle$. The latter was introduced in the main text in order to probe the correlations among the ${\bm k}$ Fourier components of the magnetization profile on two equivalent subsystems $A_i$ and $A_j$ of linear size $l_A$, as a function of their spatial separation $r_{ij}$ -- see Fig.~\ref{f.Ckr} for a sketch. 
 
  Along the same line as the calculations discussed in Appendix \ref{sec_relaxation_fluctuations_subsystem}, one finds that
\be 
 C^{zz}_{\bm k}(A_i, A_j) = \frac{1}{N} \sum_{\bm q} e^{-i{\bm q}\cdot{\bm r}_{ij}} f_A({\bm q}) S^{zz}(\bm k+ \bm q, t) ~
\ee
where $f_A({\bm q})$ is the form factor of the two equivalent subsystems. Replacing the latter with a window function in the limit $l_A \gg 1$ (see Fig.~\ref{fig_fq}), the momentum integration delivers 

\bearr
	C^{zz}_{\bm k}(\bm r,t) &\approx &\frac{f_A(\bm k)}{2} \left\lbrace 
			e^{2i\omega_{\bm k} t} \prod_{a=1}^{d} {\rm sinc} \left[
					\frac{\pi}{l_A} (2 v_{G,a}(\bm k)t - r_a )
					\right] \right.
				\nonumber \\ 
				&+&
			\left. e^{-2i\omega_{\bm k} t} \prod_{a=1}^{d} {\rm sinc} \left[
					\frac{\pi}{l_A} (2 v_{G,a}(\bm k)t + r_a )
					\right]
			\right\rbrace
		\label{eq_Ck_approx}~.~~~
\eearr
The product of the ${\rm sinc}$ functions implies that $C^{zz}_{\bm k}(\bm r,t)$ is peaked around $\bm r = \pm 2\bm v_G(\bm k) t$, namely it exhibits correlation fronts moving ballistically in space at  $\pm 2\bm v_G(\bm k)$. This result agrees with the physical intuition that correlations at wavevector $\bm k$ between two subsystems appear when they are reached by a pair of quasiparticles of wavevector $\bm k$, emitted halfway between them, and counter-propagating ballistically at speeds $\pm \bm v_G(\bm k)$.

 This result shows that the dynamics of the ${\bm k}$-dependent correlations among two extended subsystems exhibits a \emph{multiple light-cone structure}, related to the dispersion of group velocities. Each correlator $C^{zz}_{\bm k}(\bm r,t)$ possesses a different light-cone dynamics, and the ``causal cone" of correlations spreads only along the vector direction of the group velocity. All these aspects are captured by Fig.~\ref{fig_Ck}, showing moreover that, for $z<1$,  the ${\bm k}$-dependent light cone becomes ``curved" for $k=0$, revealing the nonlinear dispersion relation $\omega \sim k^z$; and moreover the correlation fronts acquire an approximate spherically symmetric structure in the long-time limit.

 \section{Time dependent DMRG simulations}
\label{dmrg_det}
In this section we provide more details about the time-dependent simulations based on the density-matrix renormalization group (DMRG). The time-dependent DMRG \cite{DegliEspostiBoschi2004,WhiteFeiguin2004,DKSV2004,Vidal2004} is an efficient algorithm to study the static and dynamical properties of quantum systems in one dimension.
Our calculations have been performed for a chain of $L=30$ sites, taking as initial state for the evolution the ground state of the fully connected Hamiltonian ($\alpha=0$):

\be 
	{\cal H}_{0XX} = -J \sum_{i,j} \left ( S_i^x S_j^x + S_i^y S_j^y \right )~.
\label{H_XX_0}
\ee
which has quantum numbers $S_{\rm tot} = N/2$ and $S^z_{\rm tot} = 0$ for the collective spin operator. 
This state is a projection onto the $S^z_{\rm tot} = 0$ sector of Hilbert space of the perfectly polarized state $ | \uparrow_x \rangle^{\otimes N}$ used in the LSW calculations.
This choice facilitates significantly the DMRG calculations, allowing us to restrict the time evolution to the $S^z=0$ sector.
We then studied the unitary dynamics generated by the Hamiltonian ${\cal H}_{\alpha{\rm XX}}$ for several values of $\alpha$ in the range $[1.5,2.5]$.
The time-evolution is based on a Runge-Kutta 4th order scheme, with a time step of $\delta t/J=0.05$. We used a variable number of states (up to 800) in order to keep the truncation error smaller than $10^{-5}$ at each time step.

\end{document}